\documentclass[twocolumn,tighten]{aastex61}
\usepackage{savesym}
\savesymbol{tablenum}
\usepackage[T1]{fontenc}
\usepackage{inputenc}
\usepackage[range-units = brackets,tophrase={-},seperr,repeatunits=false]{siunitx}
\restoresymbol{SIX}{tablenum}

\submitjournal{ApJ}

\shorttitle{Merging and feedback in NGC\,741}
\shortauthors{Schellenberger et al.}

\begin{document}
	
	\title{NGC\,741 -- Mergers and AGN feedback on galaxy group scale}

	\author[0000-0002-4962-0740]{G. Schellenberger}
	\correspondingauthor{G. Schellenberger}
	\affil{Harvard-Smithsonian Center for Astrophysics, 60 Garden Street, Cambridge, MA 02138, USA}
	\email{gerrit.schellenberger@cfa.harvard.edu}

	\author{J. M. Vrtilek}
	\affiliation{Harvard-Smithsonian Center for Astrophysics, 60 Garden Street, Cambridge, MA 02138, USA}
	
	\author{L. David}
	\affiliation{Harvard-Smithsonian Center for Astrophysics, 60 Garden Street, Cambridge, MA 02138, USA}
	
	\author{E. O'Sullivan}
	\affiliation{Harvard-Smithsonian Center for Astrophysics, 60 Garden Street, Cambridge, MA 02138, USA}
	
	\author{S. Giacintucci}
	\affiliation{Naval Research Laboratory, 4555 Overlook Avenue SW, Code 7213, Washington, DC 20375, USA}
	
	\author{M. Johnston-Hollitt}
	\affiliation{School of Chemical \& Physical Sciences, Victoria University of Wellington, Wellington, 6140, New Zealand}
	\affiliation{Peripety Scientific Ltd., PO Box 11355, Manners Street, Wellington, 6142, New Zealand}
	
	\author{S. W. Duchesne}
	\affiliation{School of Chemical \& Physical Sciences, Victoria University of Wellington, Wellington, 6140, New Zealand}
	\affiliation{Peripety Scientific Ltd., PO Box 11355, Manners Street, Wellington, 6142, New Zealand}
		
	\author{S. Raychaudhury}
	\affiliation{Inter-University Centre for Astronomy and Astrophysics, Post Bag 4, Ganeshkhind, Pune 411007, India}
	
	\begin{abstract}
		Low mass galaxy cluster systems and groups play  an essential role in upcoming cosmological studies such as those to be carried out with eROSITA.
		Though the effects of active galactic nuclei (AGNs) and merging processes are of special importance to quantify biases like selection effects or deviations from hydrostatic equilibrium, they are poorly understood on the galaxy group scale.
		We present an analysis of recent deep Chandra and XMM-Newton integrations of NGC\,741, which provides an excellent example of a group with multiple concurrent phenomena: both an old central radio galaxy and a spectacular infalling head-tail source, strongly-bent jets, a 100\,kpc radio trail, intriguing narrow X-ray filaments, and gas sloshing features. Supported principally by X-ray and radio continuum data, we address the merging history of the group, the nature of the X-ray filaments, the extent of gas stripping from NGC\,742, the character of cavities in the group, and the roles of the central AGN and infalling galaxy in heating the intra-group medium.
	\end{abstract}
	

	\section{Introduction}\label{sec:intro}
	Galaxy clusters, as the most massive gravitationally relaxed systems in the Universe, are excellent tools to study cosmology, especially the phenomena of dark matter and dark energy. The cluster mass function in particular is a sensitive probe of cosmological parameters. X-ray emission from the hot intracluster medium (ICM) traces the most massive visible component of clusters and enables derivation of both the total gravitational mass and that of the radiating baryonic component.
	A deep understanding of the evolution of the baryon distribution in the Universe is essential for modeling the structure formation process, especially for low-mass objects where non-gravitational effects become progressively more dominant.
	Furthermore, low-mass objects like galaxy groups are ideal for tracing baryons, since about 50\% of galaxies reside in galaxy groups, while galaxy clusters host only a few percent (\citealp{1998ApJ...503..569E}).
	In the hierarchical formation process, high mass galaxy clusters are the final stage of evolution, while galaxy groups are the predominant site of merging processes.
	Although extended emission observed in galaxy groups can be approximated as a scaled-down version of typical cluster emission, the gas properties in groups show differences in both the scaling relations (\citealp{2011A&A...535A.105E,Lovisari2015}) and AGN feedback properties (\citealp{2009ApJ...693.1142S,Bharadwaj2014}); these differences contribute to the justification of the study of groups independently of clusters.
	
	Galaxy groups show a variety of populations. The ones dominated by late type galaxies are usually poorer in hot, X-ray bright gas than the groups rich in ellipticals, and sometimes contain significant amounts of cold gas.
	It is well known that interactions of spiral galaxies enrich the environment (e.g., \citealp{2010ApJ...723..197K}), leading to intergalactic HI clouds or filaments. These interactions trigger the turnover of a galaxy group being dominated by elliptical galaxies, sometimes accompanied by star formation. However, it remains unclear how the hot ICM is influenced by these processes, especially how it reaches this hot state. Although dominant sources of heat are believed to be galaxy infall and shock heating, a detailed understanding of the hot intra group medium is important for structure formation scenarios.

	XMM-Newton and Chandra observations have shown in the past that feedback from the AGN in the central dominant galaxy in groups and clusters prevents the initially claimed dramatic cooling flow (e.g., \citealp{2004ApJ...607..800B}) and is of high importance for the total energy budget for the cluster. On the other hand, the presence of star formation in these galaxies seems to indicate radiative gas cooling in the center to some extent (e.g., \citealp{1995MNRAS.276..947A,2008ApJ...687..899R}).
	Furthermore, major and minor mergers in the clusters and groups have a severe influence on the thermodynamic and chemical environment (e.g., \citealp{2010ApJ...717..908Z}). While stronger mergers can create shock fronts, ram pressure stripping and cold fronts, less energetic merging events produce disturbed gas motions or displace the cores of clusters, and will also lead to a redistribution of heavy elements in the gas.
	
	In this context, NGC\,741 is a very interesting and possibly unique system: based on a relatively short X-ray exposure,  it has previously been reported to host a ghost cavity, a region of low X-ray surface brightness evacuated by the central AGN. However, the non-thermal radio emission from the relativistic particles lies below the sensitivity of previous observations (\citealp{2008MNRAS.384.1344J}). Furthermore, it is known to host a radio bridge between the brightest cluster galaxy (BCG) and a close member galaxy, NGC\,742, as discussed in \cite{1985ApJ...291...32B}. Also, further extended radio emission around the BCG was detected, but, as we discuss in this work, it originates not from the AGN of the BCG, NGC\,741, but from NGC\,742, and marks the trajectory during the encounter of the two galaxies. The radio emission does not originate only from one double-lobed radio source, as one might at first suspect.
	
	In this work we focus on deepening the understanding of the merger history of the NGC\,741 system, including the  X-ray filaments between NGC\,742 and the BCG, NGC\,741. For this work we make use of 150\,ks of Chandra data and 80\,ks of XMM-Newton data jointly awarded in Chandra cycle 16, supplemented by high resolution, multi-frequency radio images from the Very Large Array (VLA) and Giant Metrewave Radio Telescope (GMRT). Additionally, we consider the spectral properties of the system at low radio frequencies (72-231\,MHz) using data from the Murchison Widefield Array (MWA).
	
	Throughout this paper we assume a flat $\Lambda$CDM cosmology with the following parameters:\\$\Omega_{\rm m} = \num{0.27}$, $\Omega_\Lambda = \num{0.73}$, $H_0 =  h \cdot \SI{100}{km~s^{-1}~Mpc^{-1}}$ with $h = \num{0.71}$, and uncertainties are stated at the 68\% confidence level unless stated otherwise.

	\section{Data reduction and analysis}\label{sec:data}
	For our analysis we we employed both archival X-ray data from early in the Chandra and XMM missions \textemdash{} 30ks from Chandra cycle 2 and approximately 31ks from XMM-Newton \textemdash{} as well as deeper and more recent integrations proposed jointly in Chandra cycle 16 specifically for this work \textemdash{} 150ks from Chandra and 76ks from XMM-Newton (see Table \ref{tab:data}). For the detection of spatial features and the imaging analysis we rely on the Chandra data.
	
	For our radio study, we complement GMRT observations at $\SI{235}{MHz}$ and $\SI{610}{MHz}$, first presented in \cite{2011ApJ...732...95G},
	with an analysis of archival VLA observations at $\SI{1.4}{GHz}$ and $\SI{4.8}{GHz}$ and GMRT data at $\SI{150}{MHz}$, taken as part
	of the TIFR GMRT Sky Survey (TGSS\footnote{http://tgssadr.strw.leidenuniv.nl}). We also use images and fluxes from the
	GaLactic and Extragalactic All-sky MWA (GLEAM) survey \citep{2015PASA...32...25W}, covering
	the entire sky south of declination +30$^{\circ}$  with 20 frequency bands between $\num{72}$ and $\SI{231}{MHz}$, a good sensitivity
	of $\SI{10}{mJy\,beam^{-1}}$, and an angular resolution of $\SI{100}{arcsec}$. (In the course of the radio analysis,
	discrepancies noted between the GMRT $\SI{235}{MHz}$ flux density reported by \cite{2011ApJ...732...95G} and
	the GLEAM extrapolation were resolved after reprocessing of the GMRT data: see \S2.3.)
	
	In the following we briefly describe our data reduction procedure.
	
	\subsection{Chandra}
	The three available Chandra observations (Observation ID 2223, 17198, 18718) all imaged NGC\,741 on the ACIS-S3 chip and exhibit a total raw exposure time of $\SI{180}{ks}$.
	We follow the standard data reduction task using the CIAO software package version 4.8 including the contributed scripts. After creating new event files using the latest CIAO calibration of March 2016 (CALDB 4.7.1) using the \verb|chandra_repro| task, we perform a lightcurve cleaning to detect flares in the data.
	We did not detect any bad time intervals, so the final exposure times are $\SI{30}{ks}$, $\SI{91}{ks}$ and $\SI{59}{ks}$, respectively for observation ID 2223, 17198 and 18718.
	For the spectral analysis we use the \verb|xspec| software version 12.9.0o included in Heasoft 6.18. This includes the updates\footnote{\url{https://heasarc.gsfc.nasa.gov/docs/xanadu/xspec/issues/issues.html}} to \verb|xspec| to correct for the underestimation of apec normalization and abundance by a factor of  $1+z$.
	The spectral background was taken into account by using the provided blank sky observations, with a normalization factor scaled by the count rate at high energies, $\SIrange{9.5}{12}{keV}$.
	For outer low surface brightness regions we use the stowed events files (recording the particle background) and simultaneously fit the RASS spectra to them\footnote{\url{http://heasarc.gsfc.nasa.gov/cgi-bin/Tools/xraybg/xraybg.pl}}.

	\subsection{XMM-Newton}
	To reduce the XMM-Newton EPIC observation (0748190101) we used the default tasks \verb|emchain| and \verb|epchain| within the SAS software package (version 15.0.0).
	The observation was split into two parts of $\SI{3.5}{ks}$ and $\SI{71}{ks}$ exposure time in the MOS data, but for the following procedure we ignored the short part.
	For MOS data we included all events with \verb|PATTERN <= 12| and flagged \verb|#XMMEA_EM|, while for PN data we included only \verb|PATTERN <=4| and \verb|FLAG==0|. The PN data were corrected for out-of-time events. The lightcurve cleaning was performed using the \verb|ESAS| tasks \verb|mos-filter| and \verb|pn-filter|, resulting in a 60\% reduction in useful exposure time.
	The spectral background was modeled following the procedure described in \cite{2015A&A...575A..37M}.
	
	\begin{deluxetable}{cccc}
		\tablecaption{X-ray and optical data for NGC\,741 \label{tab:data}}
		\tablehead{
			\colhead{Instrument} & \colhead{Observation} & \colhead{exposure} & \colhead{resolution}
		}
		\startdata
		Chandra/ACIS-S & 2223 & $\SI{30}{ks}$ & $\SI{0.5}{\arcsec}$ \\
		Chandra/ACIS-S & 17198  & $\SI{91}{ks}$ &$\SI{0.5}{\arcsec}$ \\
		Chandra/ACIS-S & 18718 & $\SI{59}{ks}$ & $\SI{0.5}{\arcsec}$\\
		XMM/EPIC & 0748190101 & $\SI{30}{ks}$ &  $\SI{10}{\arcsec}$ \\
		HST PC  & 6587 &  $\SI{4.2}{ks}$ & $\SI{0.05}{\arcsec}$ \\
		\enddata
		\tablecomments{Column 3 refers to the X-ray exposure time remaining after lightcurve cleaning (only the XMM exposure time was significantly reduced.}
	\end{deluxetable}

	\begin{deluxetable*}{cccccccc}
		\tablecaption{Summary of the radio observations \label{tab:radiodata}}
		\tablehead{
			\colhead{Array} & \colhead{Project} & \colhead{Frequency} & \colhead{Bandwidth} & \colhead{Date} & \colhead{Time} & \colhead{FWHM} & \colhead{r.m.s} \\
			\colhead{}      & \colhead{}  &  \colhead{(GHz)} & \colhead{(MHz)} & \colhead{} & \colhead{(min)} & \colhead{(\si{\arcsec} $\times$ \si{\arcsec})} & \colhead{(mJy beam$^{-1}$)} \\
		}
		\startdata
		GMRT       & $19_{\_}043^{a}$ (R07D37) & 0.15 &  16 & 2010 Oct 30 & 15  & $25\times 25$ & 6.5 \\
		GMRT       & 12SGA01                 & 0.24 &   8 & 2007 Aug 31 & 140 & $14\times 9$  & 0.3 \\
		GMRT       & 10SGA01                 & 0.61 &  32 & 2006 Aug 27 & 140 & $7\times 4 $  & 0.05 \\
		VLA -- DnC & AH276 & 1.4 & 12.5      & 1988 May 19 & 112 & $23\times 16$ & 0.06 \\
		VLA -- D   & AB593 & 4.8 & 12.5      & 1991 May 24 &  43 & $23\times 16$ & 0.05 \\
		VLA -- BnA & AB593 & 4.8 & 12.5      & 1991 Dec 16 &  39 & $2\times 1$   & 0.03 \\
		\enddata
		\tablecomments{Column 1: Radio telescope. Column 2: project code. Columns 3--6: observing frequency, bandwidth, date and total time. Column 7: full-width half maximum (FWHM) of the array (obtained for ROBUST=0 in IMAGR). Column 8: image r.m.s. level ($1\sigma$).}
	\end{deluxetable*}
	
	\subsection{GMRT data}
	We reduced an archival TGSS pointing at 150 MHz (R07D37) observed at 2h0m0s +05d48m00s and
	containing NGC\,741, and re-processed the 235 MHz and 610 MHz observations from
	\cite{2011ApJ...732...95G}. Details are summarized in Table \ref{tab:radiodata}.
	The data were calibrated and reduced using the NRAO\footnote{National Radio Astronomy
		Observatory.} Astronomical Image Processing System (AIPS). All data
	were collected in spectral-line observing mode using the {\em GMRT}
	hardware backend. Data affected by radio frequency interference (RFI)
	were excised using RFLAG, followed by manual flagging to remove residual
	bad data. Gain and bandpass calibrations were applied using the primary
	calibrators 3C286 and 3C147 at 150 MHz and 3C48 and 3C147 at 235 MHz and 610 MHz.
	The \cite{2012MNRAS.423L..30S} scale was used to set their flux densities at 150 MHz
	and 235 MHz; the VLA 1999.2 coefficients in SETJY were used at 610 MHz.
	Phase calibrators, observed several times during each observation, were used
	to calibrate the data in phase. A number of phase self-calibration cycles,
	followed by a final self-calibration step in amplitude, were applied to the
	target visibilities. Non-coplanar effects were taken into account using wide-field
	imaging at all frequencies, decomposing the primary beam area into smaller facets.
	We note that the flux discrepancy between the 235 MHz flux in \cite{2011ApJ...732...95G} and the GLEAM extrapolation was found to be the result of errors driven
	by one GMRT antenna (C12) located in the compact core. Removal of this antenna
	brought the fluxes into better agreement.
	
	Table \ref{tab:radiodata} summarizes restoring beams and root mean square (r.m.s)
	noise levels ($1\sigma$) of our final GMRT images, obtained setting the Briggs
	robust weighting (ROBUST) to 0 in IMAGR (\citealp{briggsPhD}).
	Finally, we corrected the images for the GMRT primary beam
	response\footnote{\url{http://www.ncra.tifr.res.in:8081/~ngk/primarybeam/beam.html}}
	using PBCOR. Residual amplitude errors are estimated
	to be within $15\%$ at 150 MHz and $10\%$ at 327 MHz and 610 MHz.

	\subsection{VLA data}
	
	We reduced {\em VLA} continuum-mode observations of NGC\,741
	at 1.4 GHz and 4.8 GHz obtained from the VLA archive (Table \ref{tab:radiodata}).
	The data were calibrated and reduced in AIPS. The flux density
	scale was set using the VLA 1999.2 coefficients for 3C286 and 3C48 in SETJY.
	Several loops of imaging and self-calibration were applied to reduce
	the effects of residual phase and amplitude errors in the data. All images
	were made using ROBUST=0. Correction for the {\em VLA} primary beam
	attenuation was applied to the images using PBCOR. Residual amplitude errors are
	within $5\%$ at all frequencies.
	
	\subsection{MWA data}
	We use the source-finding software \textsc{duchamp} \citep{2012MNRAS.421.3242W} to measure the flux density in the MWA sub-band images. For these measurements, we use a growth threshold of $2\sigma_{\mathrm{rms}}$ which allows integration of the flux density out to that $2\sigma_{\mathrm{rms}}$ value after detection of the source. The local noise in the image is determined by use of the Background And Noise Estimation tool, \textsc{bane} \footnote{\url{https://github.com/PaulHancock/Aegean/wiki/BANE}}, which performs background calculation and the deviation from this background. Rather than calculating this using every pixel at once, \textsc{bane} uses sparse pixel grids not only to save computational time, but also to account for cases where the noise changes significantly over the image. Uncertainties in integrated flux densities are not left to \textsc{duchamp} as we do not use the built-in noise calculation tool. Uncertainties are derived from \begin{equation}\label{eq:flux_uncertainty}
	\sigma_{S_{\nu}} = \sqrt{\left( f S_{\nu} \right)^2 + \left( \sigma_{\mathrm{rms}} \sqrt{N_{\mathrm{beam}}} \right)^2} \quad \left[ \mathrm{Jy} \right] \, ,
	\end{equation}
	where $N_{\mathrm{beam}}$ is the the number of beams crossing the source, and $f$ is the additional percentage uncertainty associated with the flux density scale of the image. In the case of the MWA GLEAM images, this is 8 per cent at the declination of NGC\,741.

	\section{Results}\label{sec:results}
	
	Our main results on the group properties and its dynamical state are summarized in this section, where we look at the different wavelength regimes separately.
The X-ray emission is determined by the thermodynamical properties of the ICM, while radio observations are sensitive to the non-thermal processes in galaxy clusters, such as particle acceleration in shocks or AGNs. 

	\subsection{X-ray picture}
	The X-ray data allow us to characterize the galaxy group NGC\,741 as a whole.
	The spectral and surface brightness analysis follows the descriptions outlined in \cite{2017arXiv170505842S}. For this analysis the center of the group is chosen to be the emission peak of an $\SI{25}{arcsec}$ Gaussian-smoothed, exposure corrected image, because an emission weighted approach would shift the center clearly toward NGC\,742 due to the extended emission. Using the emission peak (RA 1:56:21, Dec +5:37:45) the center coincides with the galaxy NGC\,741, which we assume to be in the center of the mass halo, and NGC\,742 is just a lower mass intruder. For all further analyses the redshift is frozen to $\num{0.0185}$ (\citealp{1999ApJS..121..287H}) and the hydrogen column density to $\SI{5.11e20}{cm^{-2}}$ (\citealp{2013MNRAS.tmp..859W}).
	
	We investigate the morphology, perform a spectral analysis of the thermal emission of the hot ICM, and, by assuming hydrostatic equilibrium, calculate the total gravitating mass. 
The knowledge of general system properties, such as temperature and mass, is crucial to interpret irregular phenomena (e.g., mergers, shocks).
	\begin{figure*}
		\centering
		\resizebox{1.\hsize}{!}{\includegraphics[width=0.49\linewidth]{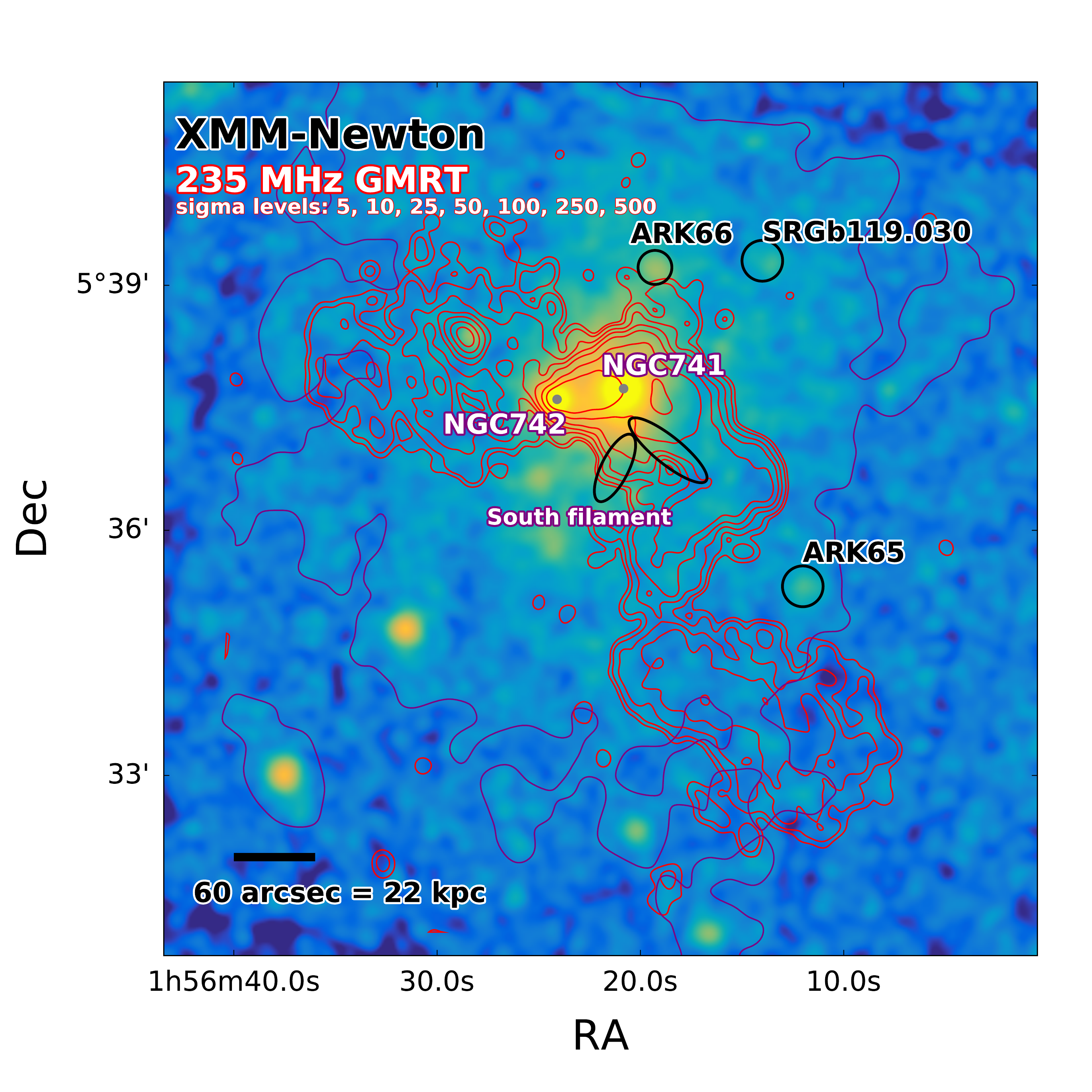} \includegraphics[width=0.49\linewidth]{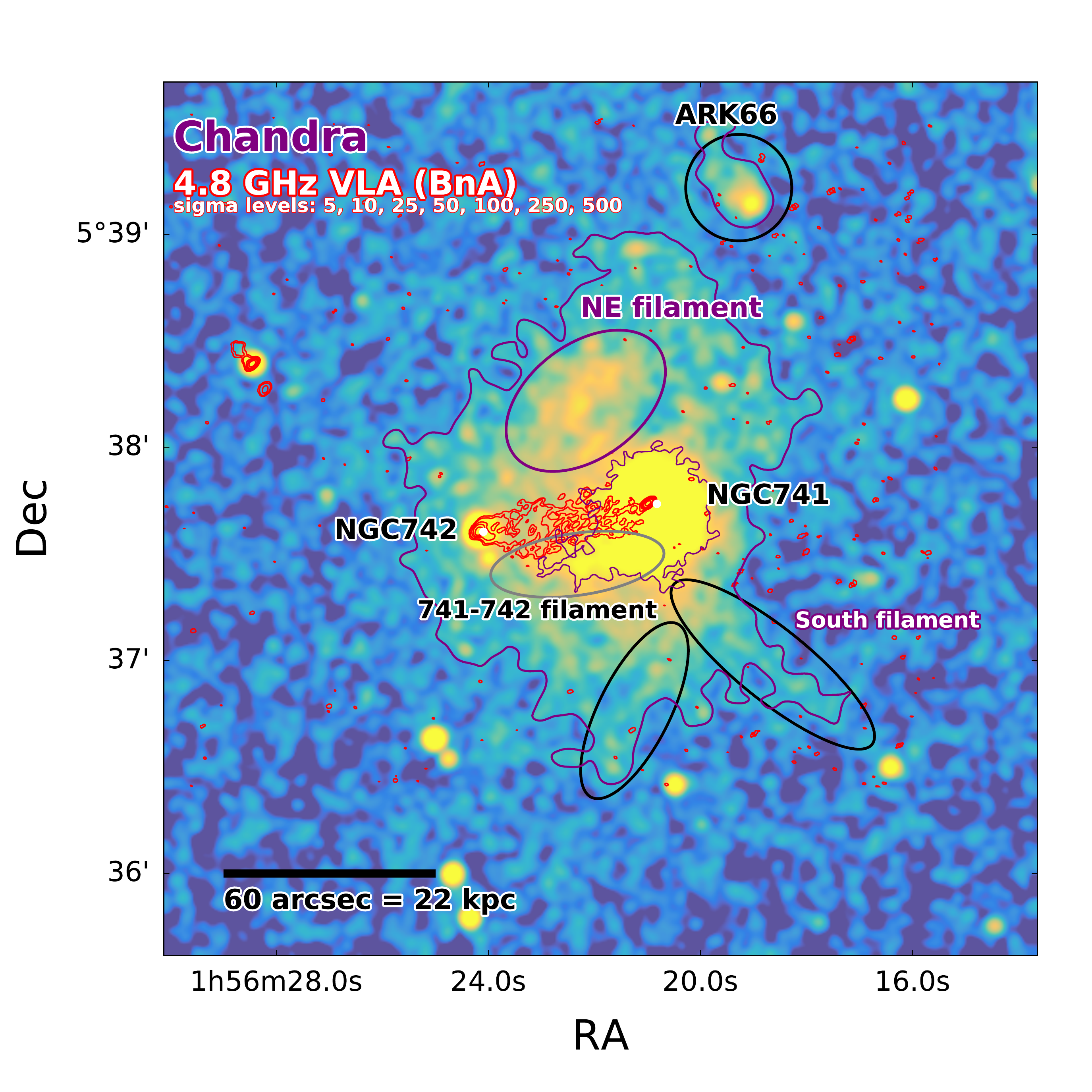}}
		\caption{Smoothed and exposure corrected image from the combined XMM-Newton/EPIC (left) and from the Chandra ACIS (right) detectors. Red contours show the radio emission detected with the GMRT at 235\,MHz (left) and with the VLA at 4.8\,GHz (right), and the purple and white contours follow the smoothed X-ray surface brightness. Angular resolution and $1\sigma$ noise values of the radio images are as listed in Table \ref{tab:radiodata}. }
		\label{fig:xmm}
	\end{figure*}
	
	\subsubsection{General X-ray morphology}
\begin{figure*}
		\centering
		\resizebox{0.4\hsize}{!}{\includegraphics{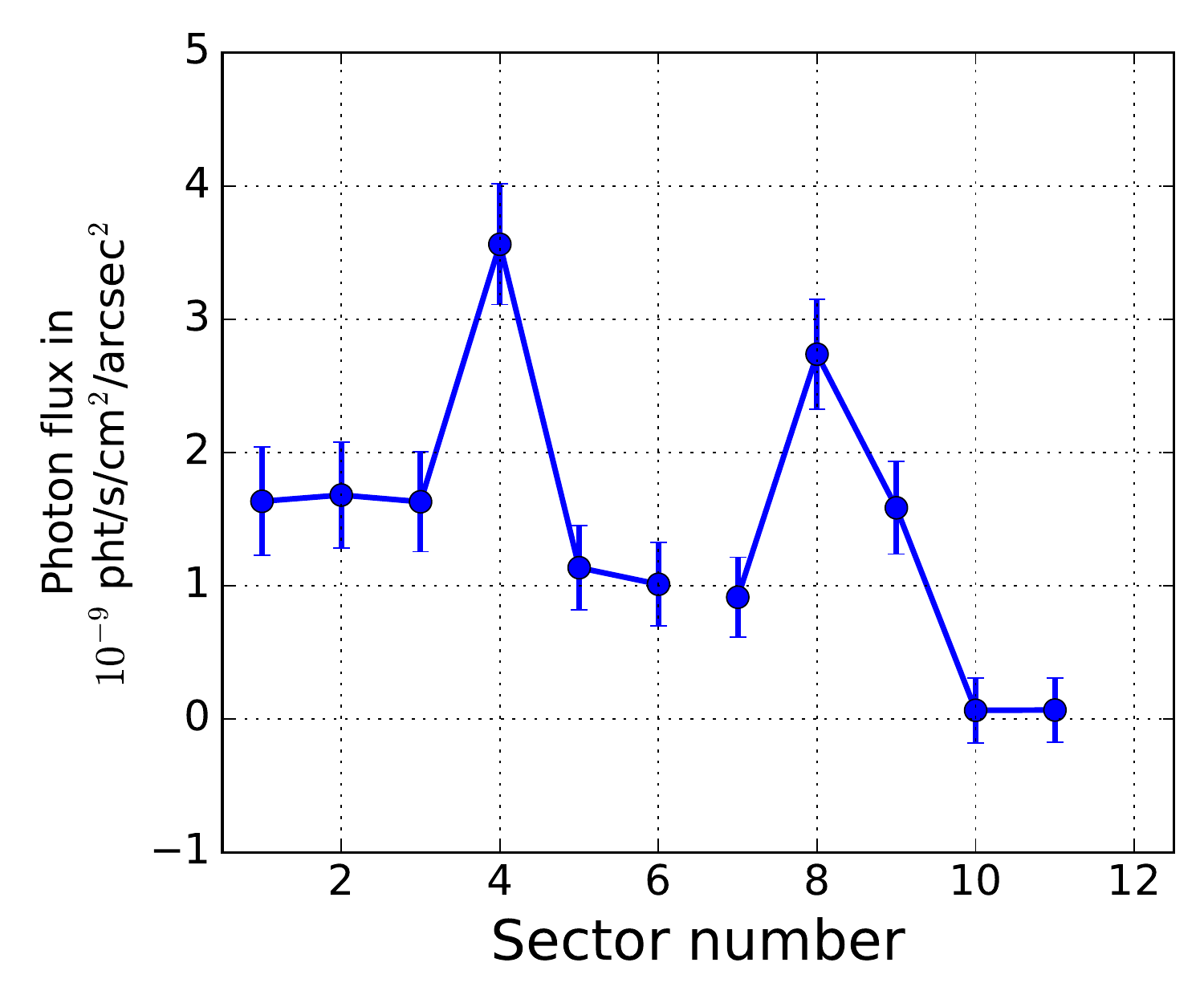}}
		\resizebox{0.4\hsize}{!}{\includegraphics[clip,trim=30px 150px 30px 150px]{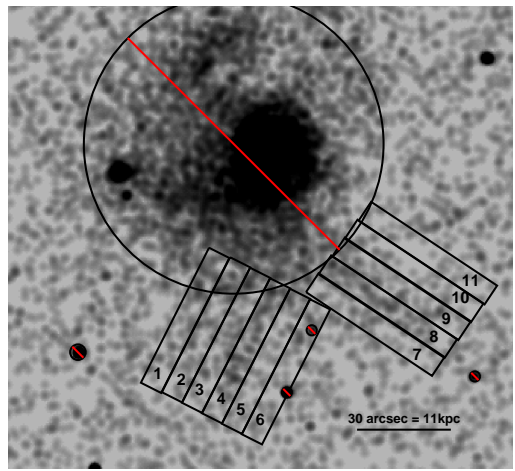}}
		\caption{Chandra surface brightness distribution in the region south to the core where two line-shaped filaments can be identified.}
		\label{fig:sbr}
	\end{figure*}

	Figure \ref{fig:xmm} shows XMM-Newton and Chandra X-ray images superposed radio contours. The radio emission arises from non-thermal synchrotron processes which lead to   a power-law spectrum (at first approximation). In this case it is the superposition of two sources, the AGN emission of NGC\,741 and NGC\,742, which leads to the given spatial distribution. On very small scales, it is even possible to trace the emission from the bent radio jets of NGC\,742.
	The left panel shows the XMM-Newton data overlaid with GMRT radio contours at $\SI{235}{MHz}$. In the core we see NGC\,741 and NGC\,742 within the irregularly shaped X-ray emission of the galaxy group, while at larger distances, three other galaxies can be identified (see Section \ref{ch:gals}). The right panel shows the Chandra data, zoomed-in on the core region, overlaid with the VLA--BnA $\SI{5}{GHz}$ contours of the radio emission between the two galaxies in the core (see also Fig. \ref{fig:bridge}). Furthermore, we are able to identify several filaments in the X-rays: The X-ray filament connection NGC\,741 and NGC\,742, a north-east filament or sub-clump and also two line-shaped features to the south of the core (see also Fig. \ref{fig:sbr}).
	To the west of NGC\,741 we notice a decrease in the X-ray surface brightness, which might be a hint of a cavity or recent merging activity.
	Although the galaxy group appears to be dynamically disturbed, there exists a very peaked surface brightness in the core indicating the presence of a cool core.

	We extract the radial surface brightness profile of the Chandra data in the $\SIrange{0.5}{2.0}{keV}$ band and correct for vignetting and variations in the exposure time using the \verb|merge_obs| task (see Fig. \ref{fig:temperatures} left bottom).
	We find a steeper surface brightness profile to the south-west when we fit several sectors separately. The excess emission to the north-east is related to an extended source, which seems to be related to the galaxy group.
	As already indicated in Fig. \ref{fig:xmm} (right) there are at least two small X-ray filaments to the south (west) of NGC\,741. We analyze those in more detail in Fig. \ref{fig:sbr}. The cylindrical shape of these objects seems remarkable, and they are almost perpendicular to each other. The one aligned to the south-west lies at the edge of the south-west radio tail.
	
	Other bright features in the residuals image are the X-ray filament between NGC\,741 and NGC\,742 ($\num{0.15} \times \SI{0.86}{arcmin}$ in size), and a $\sim 15\sigma$ decrease in surface brightness to the west of NGC\,741 (see discussion on the cavity in Sec. \ref{ch:cavity}).
	
	\subsubsection{Spectral analysis}
	For spectral extraction we use the \verb|specextract| task provided with the CIAO software package. Since we are dealing with regions of few photon counts, we use the C statistic (\citealp{1979ApJ...228..939C}) during the spectral fitting and apply a minimum number of 1 count per spectral bin. To account for the co-variance between fitting parameters (e.g., temperature and abundance) we use the Metropolis MCMC algorithm implemented in \verb|xspec|, and calculate derived quantities (like the entropy) from the chains.
	
	\begin{figure*}
		\centering
		\resizebox{1.\hsize}{!}{\includegraphics{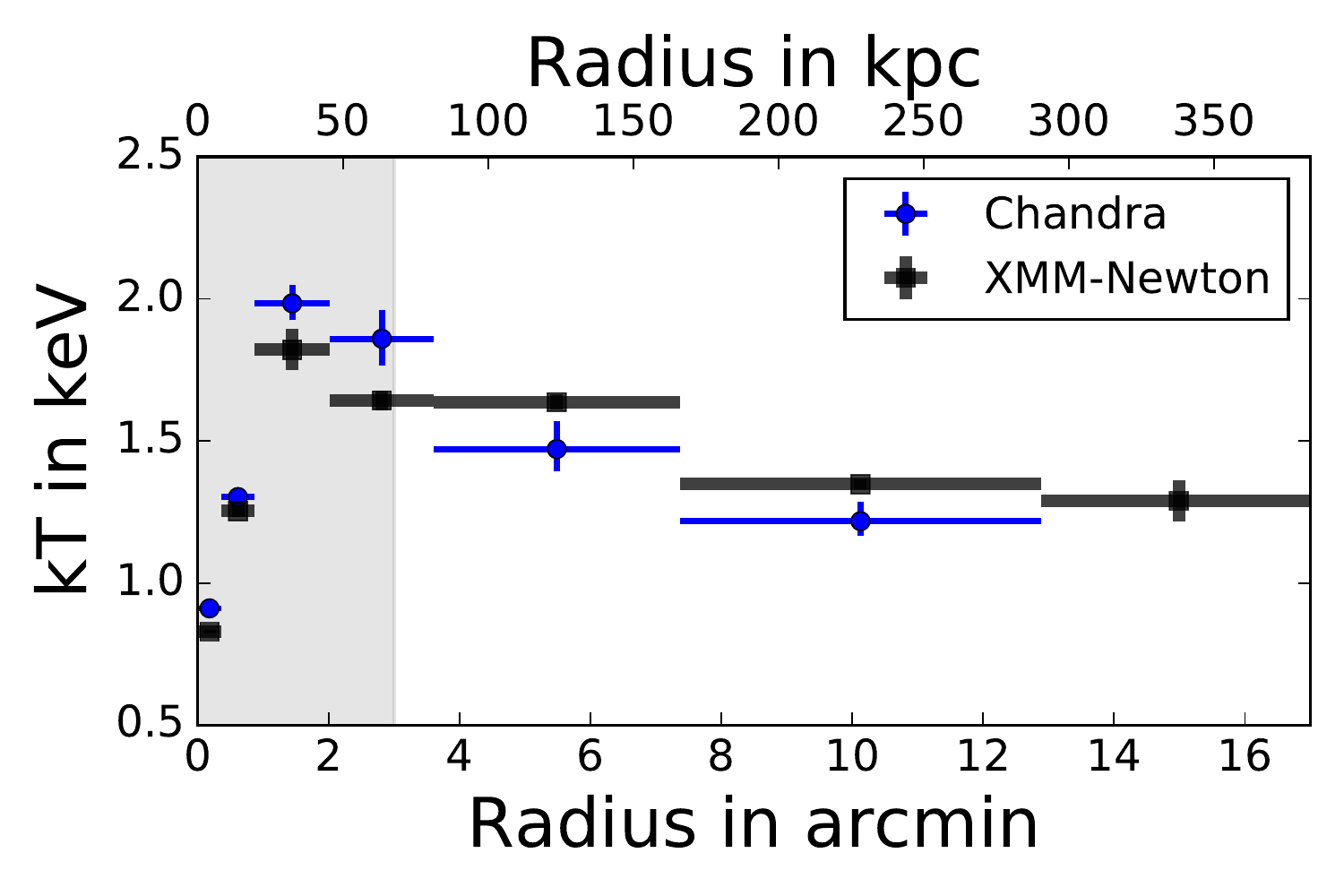}\includegraphics{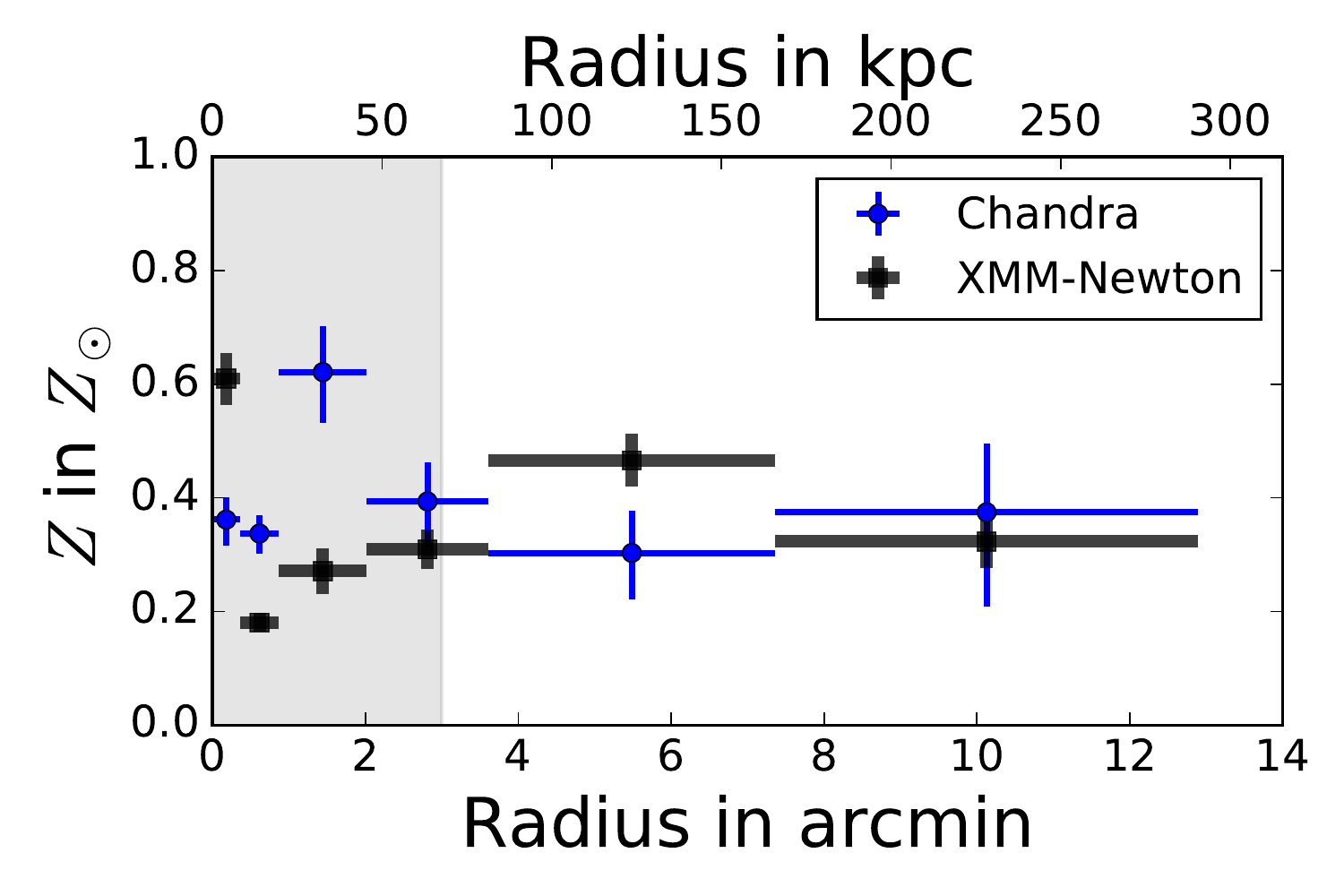}}
		\resizebox{1.\hsize}{!}{\includegraphics{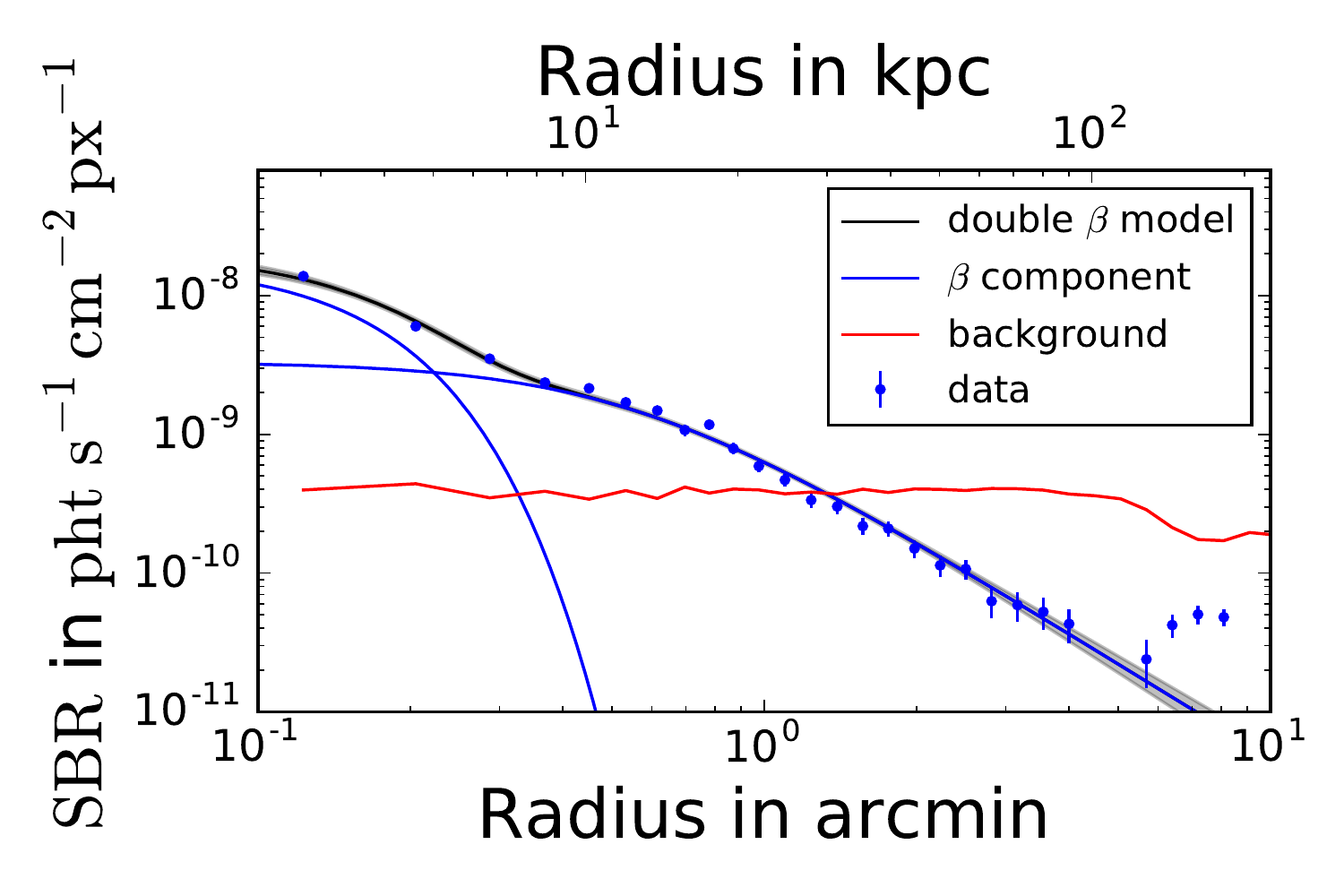}\includegraphics{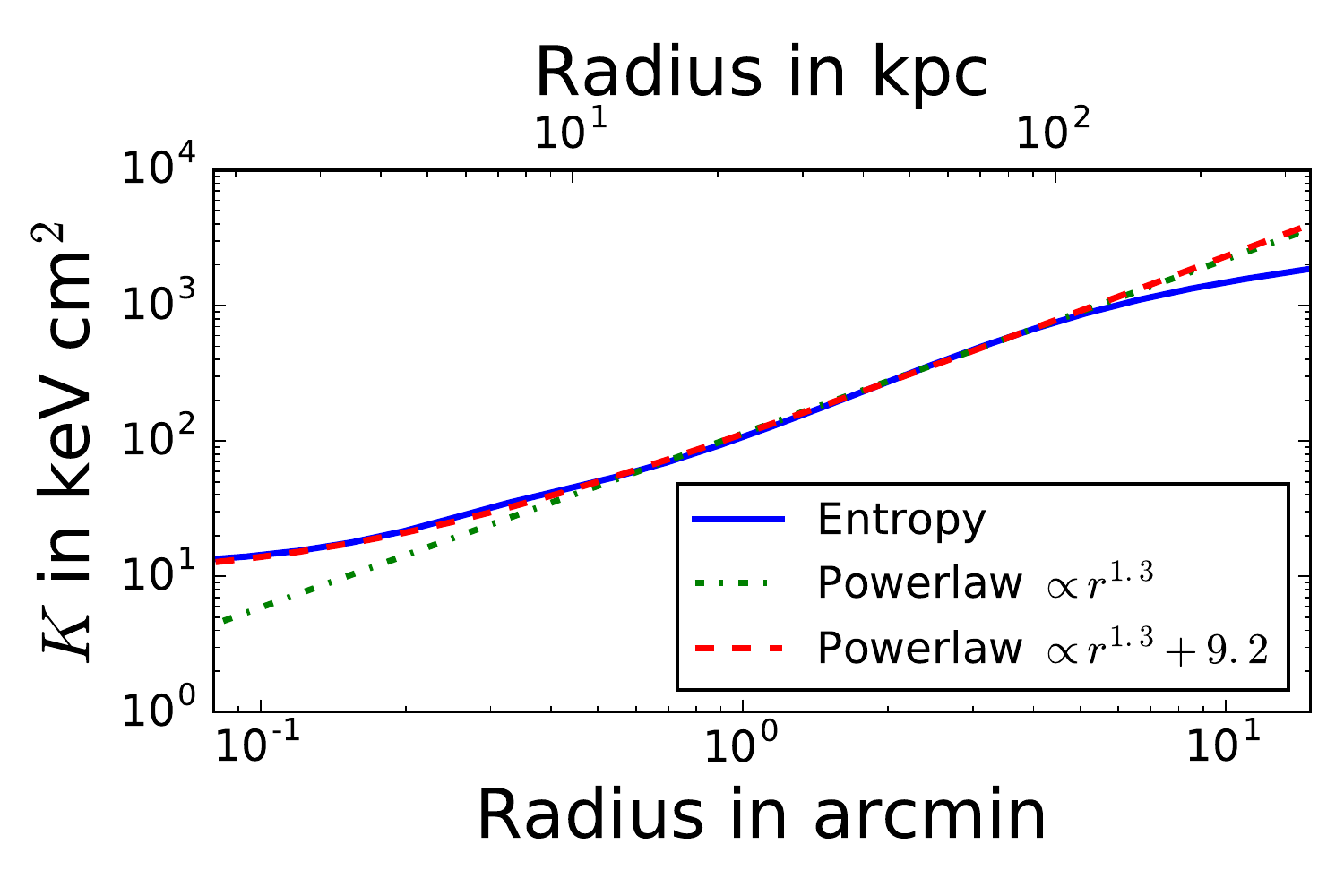}}
		\caption{\textit{Top:} Projected temperature (left) and abundance (right) profiles of the Chandra and XMM-Newton data. The gray area indicates where the Chandra FOV covers the all parts of the cluster, otherwise the extraction is along the Chandra S-chips array in NW-SE direction. \textit{Bottom:} Surface brightness (SBR, left) and entropy (right) profiles.}
		\label{fig:temperatures}
	\end{figure*}
	
	The spectral modeling of the ICM emission is described by an absorbed thermal model (\verb|phabs|$\times$\verb|apec|) using the AtomDB code version 2.0.2 (\citealp{2012ApJ...756..128F}). The relative abundance of heavy elements is set to the values of \cite{2009ARA&A..47..481A}.
	
	Figure \ref{fig:temperatures} shows the projected temperature profile extracted in annuli centered on the NGC\,741 X-ray peak and excluding point sources detected by the \verb|wavdetect| task. The galaxy group exhibits a cool core with a temperature below $\SI{1}{keV}$ in the center, and a peak temperature of around $\SI{2}{keV}$, which is unusually high for a system of this size. At larger radii the temperature drops to about $\SIrange{1.3}{1.4}{keV}$. The XMM-based profile of the relative abundance of heavy elements (Fig. \ref{fig:temperatures}) shows a sharp increase in the central bin, but this might be related to the larger PSF, since the distribution of heavy elements is complicated in this object.
		
	We compute a hydrostatic mass at a radius where the density within that radius has dropped to 500 times the critical density of the Universe ($R_{500}= \SI{560(15)}{kpc} = \SI{25.2(7)}{arcmin}$) of $M_{500} = \SI{5.2(4)e13}{M_\odot}$. The mass profile was extrapolated using an NFW model with the $c-M$ relation from \cite{2013ApJ...766...32B}. More details on this procedure are given in \cite{2017arXiv170505842S}. The gas mass at this radius is $M_\mathrm{gas, 2500} = \SI{8.6(3)e11}{M_\odot}$, which leads to a gas mass fraction of $f_\mathrm{gas, 2500} = \SI{2.9(1)}{\%}$. This value is at the lower end of what is expected for galaxy groups (\citealp{2009ApJ...693.1142S,Lovisari2015}).
	
	The entropy of the galaxy cluster ICM can be derived from the temperature and gas density,
	\begin{equation}
	K = \mathrm{k}T \cdot n^{-\frac{2}{3}}~.
	\end{equation}
	This quantity is essential to study non-gravitational effects within the ICM (see, e.g., \citealp{1999Natur.397..135P,2009ApJS..182...12C}).
	We constrain the central entropy of the core to $\SI{12(1)}{keV\,cm^{2}}$, which is typical for cool core systems.
	In a region immediately around the AGN of NGC\,742 we find $\SI{7.8(13)}{keV\,cm^{2}}$.
	In an annular region (from $\SI{6}{kpc}$ to $\SI{12}{kpc}$) we find an entropy $\SI{39(2)}{keV\,cm^{2}}$;
	although the X-ray filament connecting the AGN of NGC\,741 and NGC\,742 lies mainly in the same region as this annulus, we find an entropy of $\SI{9.9(6)}{keV\,cm^{-2}}$ in the filament, assuming it has a cylindrical shape.
	The pressures of the gas in the filament and in the annulus are in rough equilibrium, although there is a hint that the pressure in the X-ray filamentary gas is slightly higher.
	We analyzed regions in the XMM-Newton data along the south-west radio tail (see Fig. \ref{fig:xmm} left panel,  and Section \ref{ch:swtail}) to find hints for the origin of the gas. The available data do not give any conclusions for a different entropy in the regions of the south-west radio tail.
	
	\subsubsection{Temperature and metallicity maps}
\begin{figure*}
		\centering
		\resizebox{0.49\hsize}{!}{\includegraphics{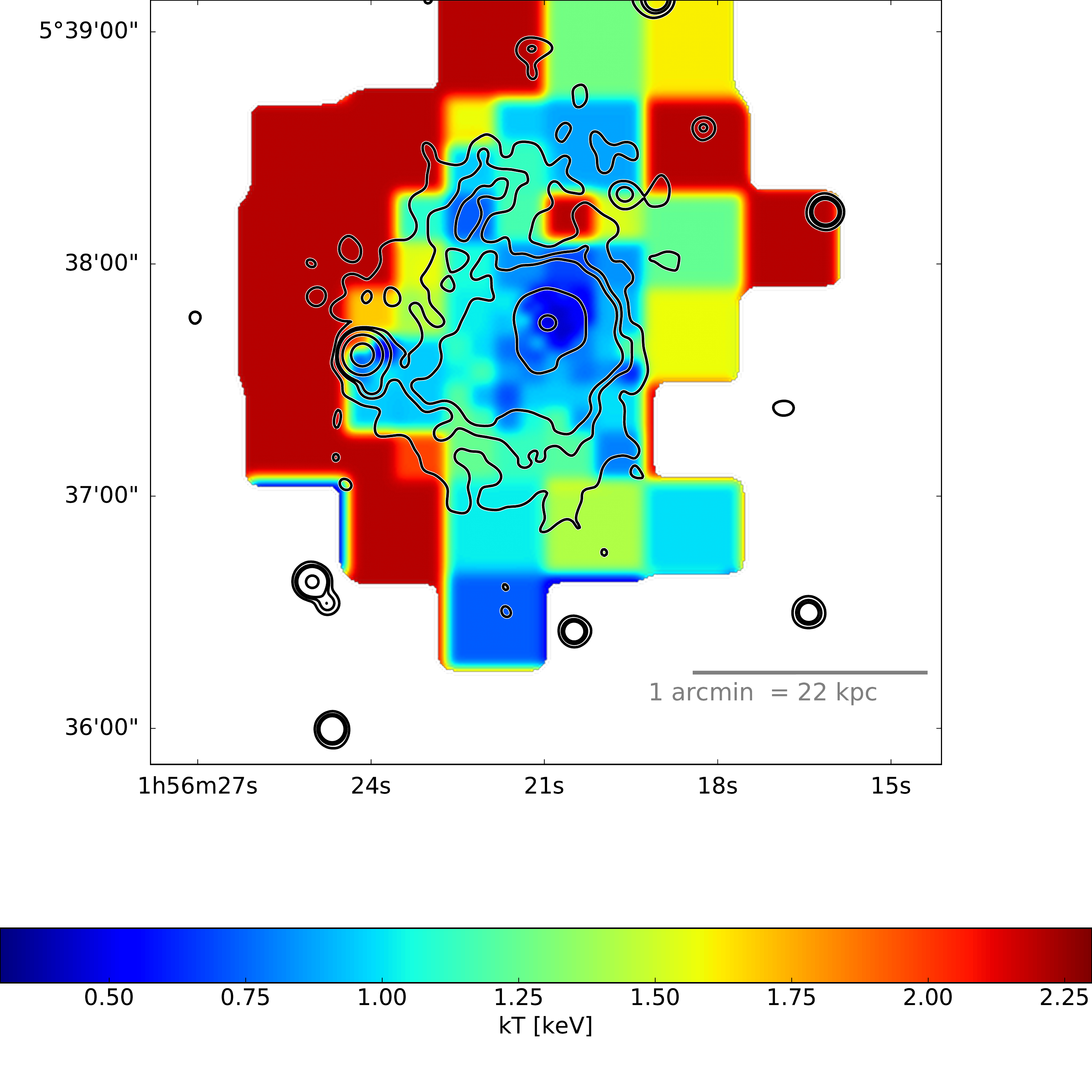}}
		\resizebox{0.49\hsize}{!}{\includegraphics{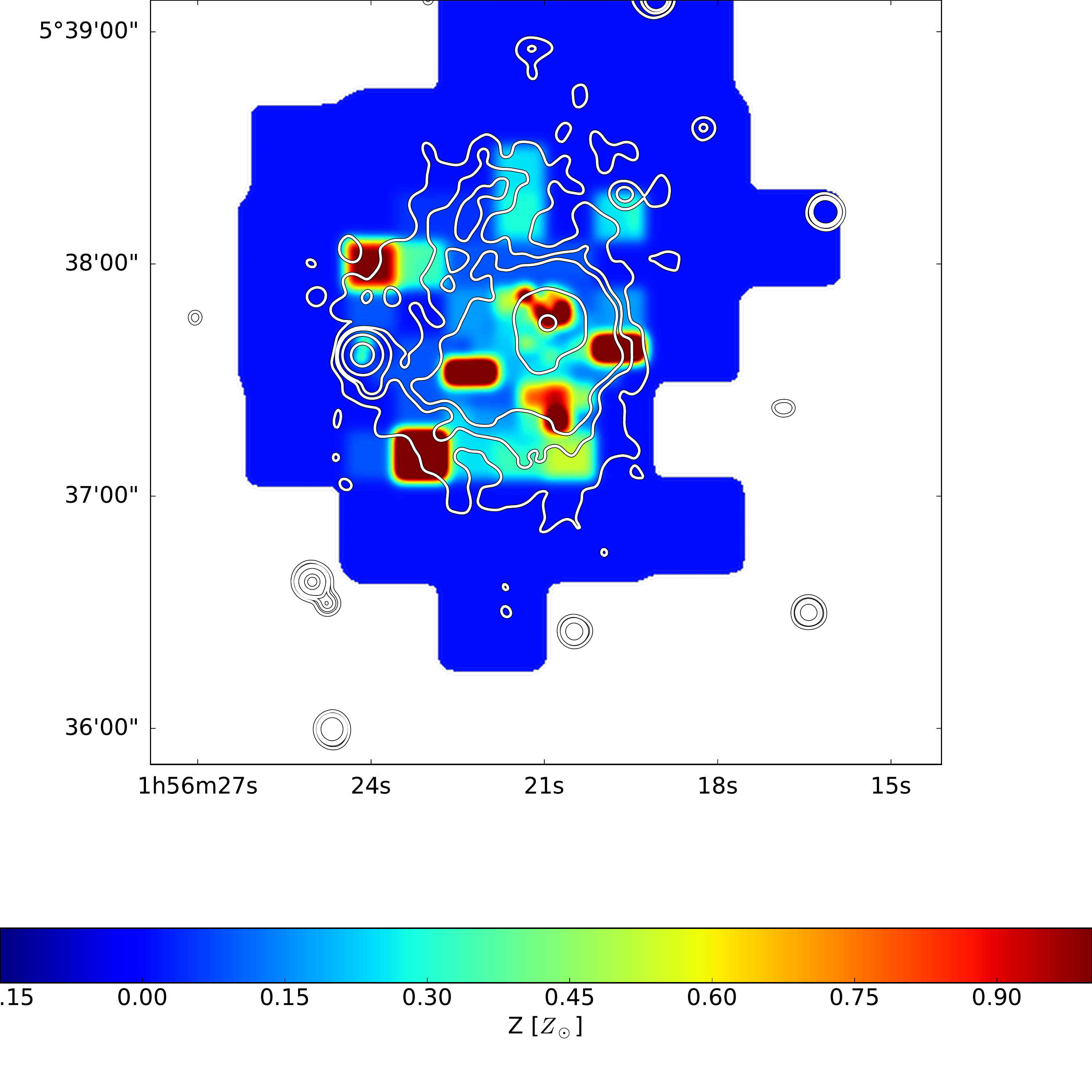}}
		\caption{Temperature and metallicity maps for NGC\,741 with the soft X-ray band contours overlaid. Only regions with a source-to-background ratio of at least 1.33 are shown. See text and Appendix \ref{app:maps} for details.}
		\label{fig:maps}
	\end{figure*}

	We construct gas temperature and metallicity maps by dividing the images into three different energy bands. Appendix \ref{app:maps} explains the details of this method. While the absolute values of temperature and metallicity might be slightly biased with this method, we are mainly interested in identifying local variations in the gas temperature and metallicity on small spatial scales.
	Figure \ref{fig:maps} shows the resulting maps for NGC\,741.
	The images are initially adaptively binned (using the CIAO task \verb|dmnautilus|) 
	to achieve a signal-to-noise of 15 in the Fe-L line (0.7-1.3~keV) and then Gaussian smoothed with $\sigma=\SI{1.5}{arcsec}$. 
	We do not consider any regions where the source count rate in the broad energy band is less than 33\% of the background count rate.
	
	For the large-scale temperature distribution, we find a similar trend as derived from our full spectroscopic analysis; however, it seems that there is an elongation of cooler gas along the north-south direction. There is also a  sudden 50\% increase in the temperature east of NGC\,742. The connection between NGC\,741 and NGC\,742 is filled mostly with cooler gas, but there is also some slightly warmer gas, indicating that the gas between the two galaxies has had enough time to mix with the surrounding, slightly hotter ICM.
	To the north-east of NGC\,741 we detect another cooler region that corresponds to an enhanced surface brightness region. 
	Finally, there is cooler gas to the south-west of NGC\,741 that corresponds to the filament seen in Fig. \ref{fig:sbr} (region 8).
	
	The abundance map shows, on average, a low metallicity of about 0.1 to 0.3 solar. These abundances are lower than those shown in 
	the azimuthally-averaged abundance profile, but the abundance map shows that there are local peaks and enhancements in the metallicity. Most of the higher abundance regions are located toward the south of NGC\,741. Indeed, we verified with a full spectroscopic analysis that these spatial fluctuations in the metallicity are real. For this analysis we used the C-statistic 
	and an MCMC approach. Contrary to the spectroscopic analysis, we find a high abundance of heavy elements in the center of NGC\,741.

	\subsubsection{Other galaxies and AGNs}
	\label{ch:gals}
	In the normalized residuals image of the surface brightness, we identify three smaller X-ray halos related to group member galaxies
	(see Fig \ref{fig:xmm}). Two of them have been mentioned before by \cite{2008ApJ...679.1162J}, ARK66 and ARK65, while SRGb119.030 has not previously been identified in the X-rays. Only ARK66 is bright enough for a spectroscopic analysis and we measure a temperature of $0.33^{+0.09}_{-0.04}\,\mathrm{keV}$ and a luminosity of $L_\mathrm{X, 0.5-2.0} = 9.4^{+1.9}_{-1.4}\times 10^{39}\,\si{erg\,s^{-1}}$. For the other two galaxies, we are able to constraint only the luminosity with a reasonable precision. For SRGb119.030 we obtain $L_\mathrm{X, 0.5-2.0} = 2.6^{+0.7}_{-0.4}\times 10^{39}\,\si{erg\,s^{-1}\,}$, and for ARK65 we obtain $L_\mathrm{X, 0.5-2.0} = 3.2^{+0.4}_{-0.4}\times 10^{39}\,\si{erg\,s^{-1}\,}$. Based on the K-band luminosities, ARK66 has an unusually high X-ray luminosity. In the case of ARK66, the X-ray emission has a head-tail structure, suggesting that the galaxy is moving toward the south-west. This head-tail structure is likely to be gas that has been ram-pressure stripped from the galaxy (also see recently discovered X-ray tails associated with other galaxies, e.g., \citealp{2006ApJ...637L..81S,2015arXiv151003708S}). From measured redshifts by  \cite{2004ApJ...607..202M,2011MNRAS.416.2840L}, we find a line-of-sight velocity difference between ARK66 and NGC\,741 of $\SI{772(9)}{km\,s^{-1}}$. Since we see an extended tail behind ARK66, the trajectory of ARK66 cannot be exactly along the line-of-sight. If we assume a $45^\circ$ orbital inclination relative to the plane of the sky, then the 3D velocity must be about $\SI{1100}{km\,s^{-1}}$, which is relatively high given the projected distance from the group center of $\SI{32}{kpc}$.
	
	A comparison of the X-ray fluxes for the AGN in the centers of each galaxy (NGC\,741 and NGC\,742), shows that there is no significant variation over almost 15 years for NGC\,741, but there is a 50\% flux increase for the AGN in NGC\,742. The current flux of the AGN in NGC\,742 is $\SI{1.30(4)e-13}{erg\,s^{-1}\,cm^{-2}}$, while the flux of the AGN in NGC\,741 is about two orders of magnitude lower.
	
	\subsection{Radio picture}
	\begin{figure}
		\centering
		\resizebox{0.95\hsize}{!}{\includegraphics[clip,trim=0px 100px 0px 100px,width=0.95\linewidth]{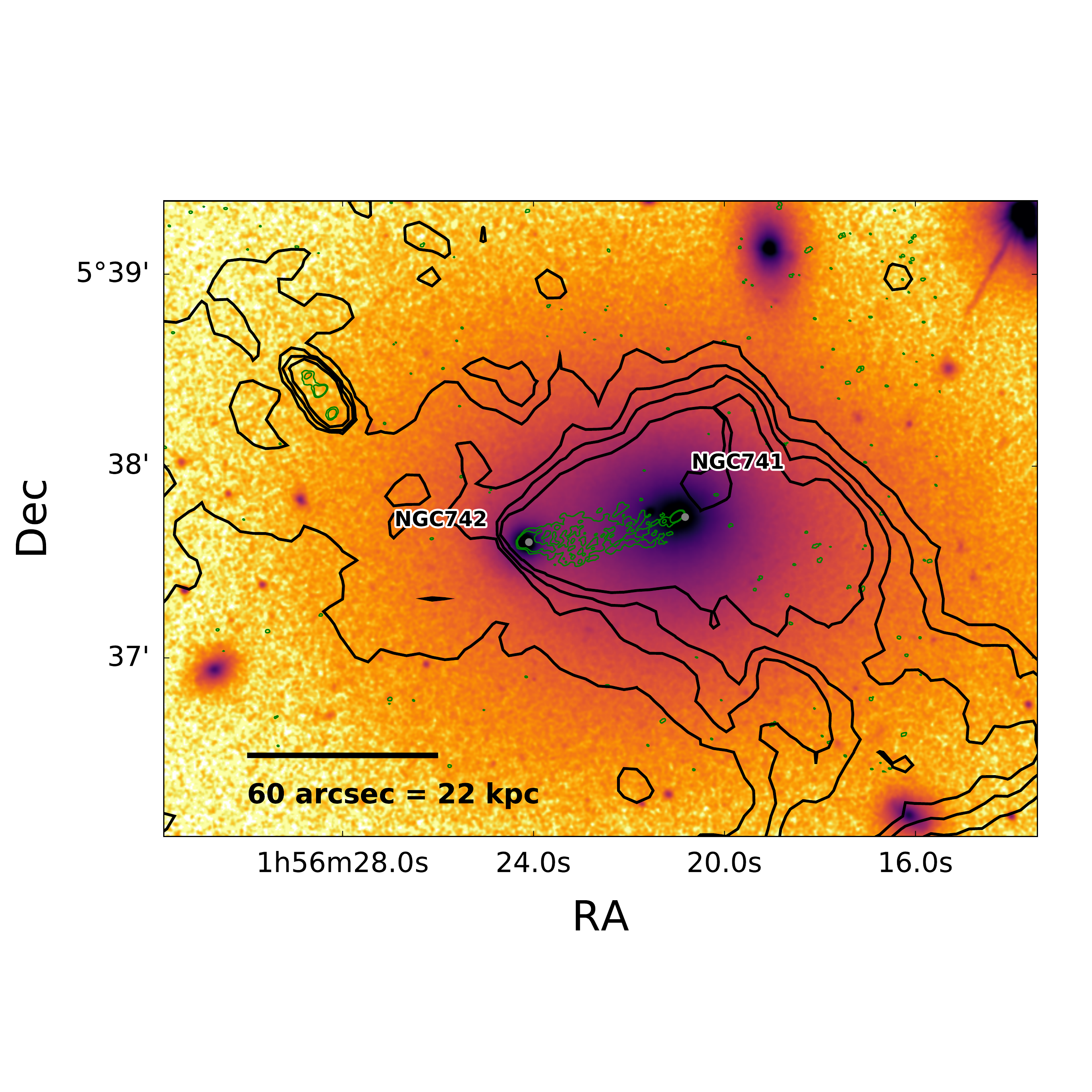}}
		\caption{SDSS r-band image with radio contours overlaid: In black $\SI{610}{MHz}$ GMRT contours ($5,25,50,100 \sigma$), and  VLA $\SI{5}{GHz}$ contours in green ($5,13 \sigma$).}
		\label{fig:opticalradio}
	\end{figure}
	The radio emission traces the synchrotron emission of high energy particles that were ejected by the AGN.
	By analyzing the radio fluxes in different bands, we can draw conclusions on the emitted energy, age of the particles, and 
	magnetic field strength. We have data available from the GMRT, VLA and the MWA on NGC 741.
	
	We initially identify extended emission and point sources at all frequencies. The broad, extended emission can be seen in Fig. \ref{fig:xmm} with the X-ray emission and Fig. \ref{fig:opticalradio} on top of the optical SDSS data. The radio emission
	to the east of NGC\,741 extends approximately $\SI{80}{kpc}$ from the central AGN, while the radio tail to the south-west has a maximum extent of $\SI{110}{kpc}$.
	
	\subsubsection{The bent jets of NGC742}
	\begin{figure}
		\centering
		\resizebox{0.9\hsize}{!}{\includegraphics[width=0.9\linewidth]{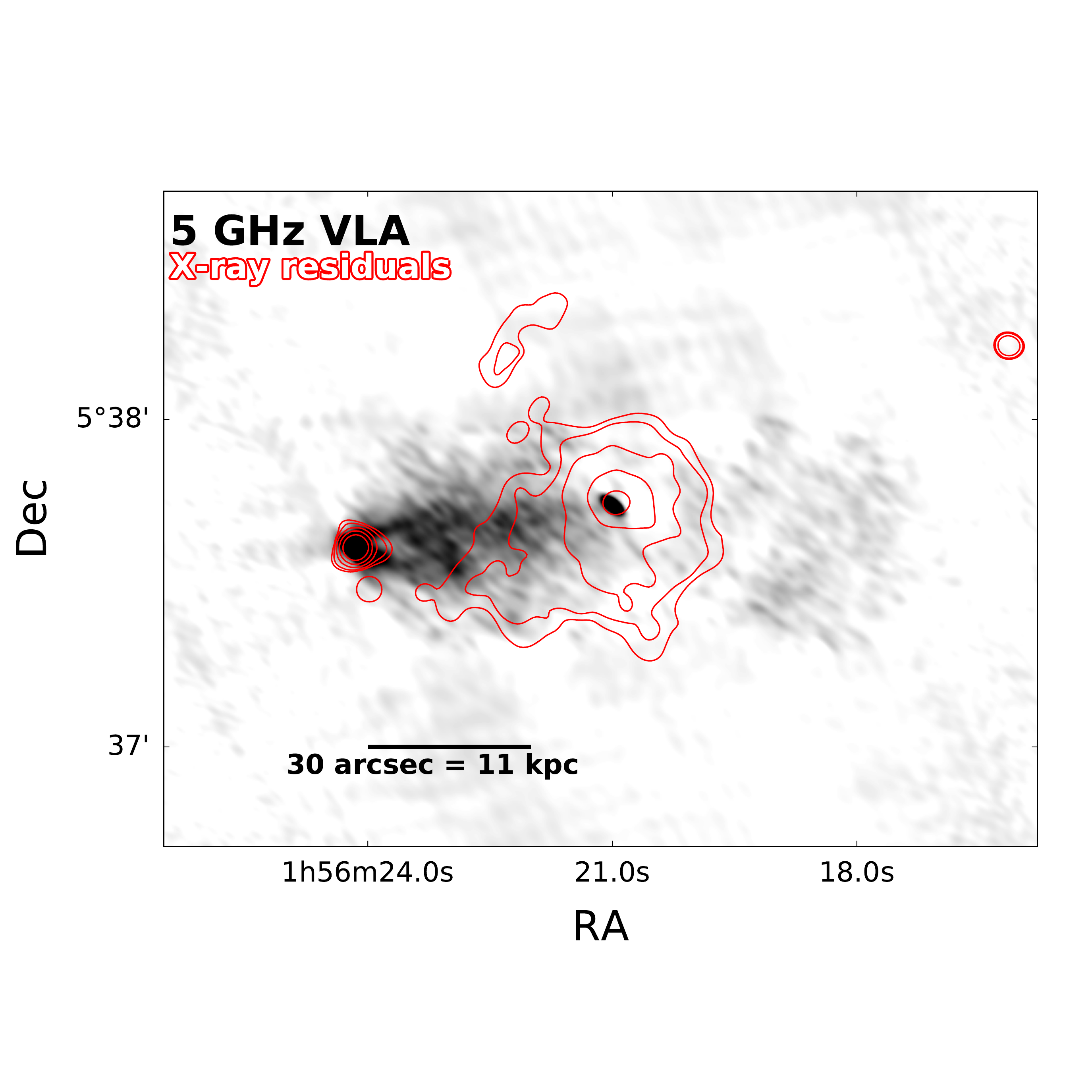}}
		\caption{Radio jets of NGC742, formerly classified as bridge connecting NGC\,742 and NGC\,741.}
		\label{fig:bridge}
	\end{figure}
	
	The high resolution image at $\SI{5}{GHz}$ with a beamsize of $\num{2} \times \SI{1}{arcsec}$ (see Fig. \ref{fig:bridge}), provides a more detailed view of the radio emission between NGC\,742 and NGC\,741. NGC\,742 appears to be a head-tail radio galaxy with strongly bent jets to the west. 
	The emission is brightest around NGC\,742 and becomes fainter toward the west. We also see the v-shaped structure of the jet around NGC\,742, typical for bent radio jets.
	The radio surface brightness decreases toward NGC\,741, beyond which we do not detect small scale radio emission. NGC\,741 itself does not show any radio jets and is only visible as a point source. From a high resolution VLA observation at $\SI{1.4}{GHz}$,
	we detect a clear spectral steepening from NGC\,742 to NGC\,741, with spectral indices\footnote{we define the spectral index $\alpha$ as $S_\nu \propto \nu^\alpha$, where $S_\nu$ is the flux density at frequency $\nu$} between -0.5 and -1 (M. Birkinshaw, private communication).
	In Section \ref{ch:ngc742merging} we present a discussion concerning the shape of the bent jets.

	\subsubsection{Radio spectrum}
	\label{ch:flux_comp}
	
	To compare the radio emission of NGC\,741 with other telescopes, 
	we need to compute the integrated flux. 
	We calculate integrated fluxes and uncertainties in our high resolution images from VLA and GMRT using
	the \verb|radioflux| python tool by M. Hardcastle\footnote{\url{www.extragalactic.info/~mjh/radio-flux.html}}.
	In addition, data from the MWA GLEAM survey was measured using the Duchamp software package (\citealp{2012MNRAS.421.3242W}).
	
	\begin{deluxetable}{ccc}
		\tabletypesize{\footnotesize}
		\tablecaption{Radio fluxes for NGC\,741 \label{tab:rflux}}
		\tablehead{
			\colhead{Instrument} & \colhead{Frequency} & \colhead{Flux} \\
			\colhead{} & \colhead{$\si{MHz}$} & \colhead{$\si{Jy}$}
		}
		\startdata
		VLSS-R    & 74 & $\num{10.36(79)}$\\
		MWA/GLEAM & 76   & $\num{8.58(72)}$ \\
		MWA/GLEAM & 84   & $\num{7.45(62)}$ \\
		MWA/GLEAM & 92   & $\num{6.78(57)}$ \\
		MWA/GLEAM & 99   & $\num{6.22(52)}$ \\
		MWA/GLEAM & 107   & $\num{6.25(52)}$ \\
		MWA/GLEAM & 115   & $\num{6.13(50)}$ \\
		MWA/GLEAM & 122   & $\num{5.54(45)}$ \\
		MWA/GLEAM & 130   & $\num{5.37(44)}$ \\
		MWA/GLEAM & 143   & $\num{5.21(43)}$ \\
		MWA/GLEAM & 150   & $\num{4.89(40)}$ \\
		MWA/GLEAM & 158   & $\num{4.67(38)}$ \\
		MWA/GLEAM & 166   & $\num{4.51(37)}$ \\
		MWA/GLEAM & 173   & $\num{4.34(36)}$ \\
		MWA/GLEAM & 181   & $\num{4.09(34)}$ \\
		MWA/GLEAM & 189   & $\num{4.04(33)}$ \\
		MWA/GLEAM & 196   & $\num{3.76(31)}$ \\
		MWA/GLEAM & 204   & $\num{3.68(31)}$ \\
		MWA/GLEAM & 212   & $\num{3.57(30)}$ \\
		MWA/GLEAM & 220   & $\num{3.47(29)}$ \\
		MWA/GLEAM & 227   & $\num{3.53(30)}$ \\
		GMRT      & 150 & $\num{5.44(82)}$ \\
		GMRT      & 240 & $\num{4.25(43)}$ \\
		GMRT      & 613 & $\num{1.61(16)}$ \\
		NVSS      & 1400& $\num{1.04(5)}$  \\
		VLA (DnC) & 1411 & $\num{1.02(5)}$ \\
		VLA (DnC) & 4841 & $\num{0.30(2)}$ \\
		Becker+ 1991 & 4850 & $\num{0.31(3)}$\\
		\enddata
	\end{deluxetable}
	
\begin{figure*}
		\centering
		\resizebox{0.9\hsize}{!}{\includegraphics[width=0.9\linewidth]{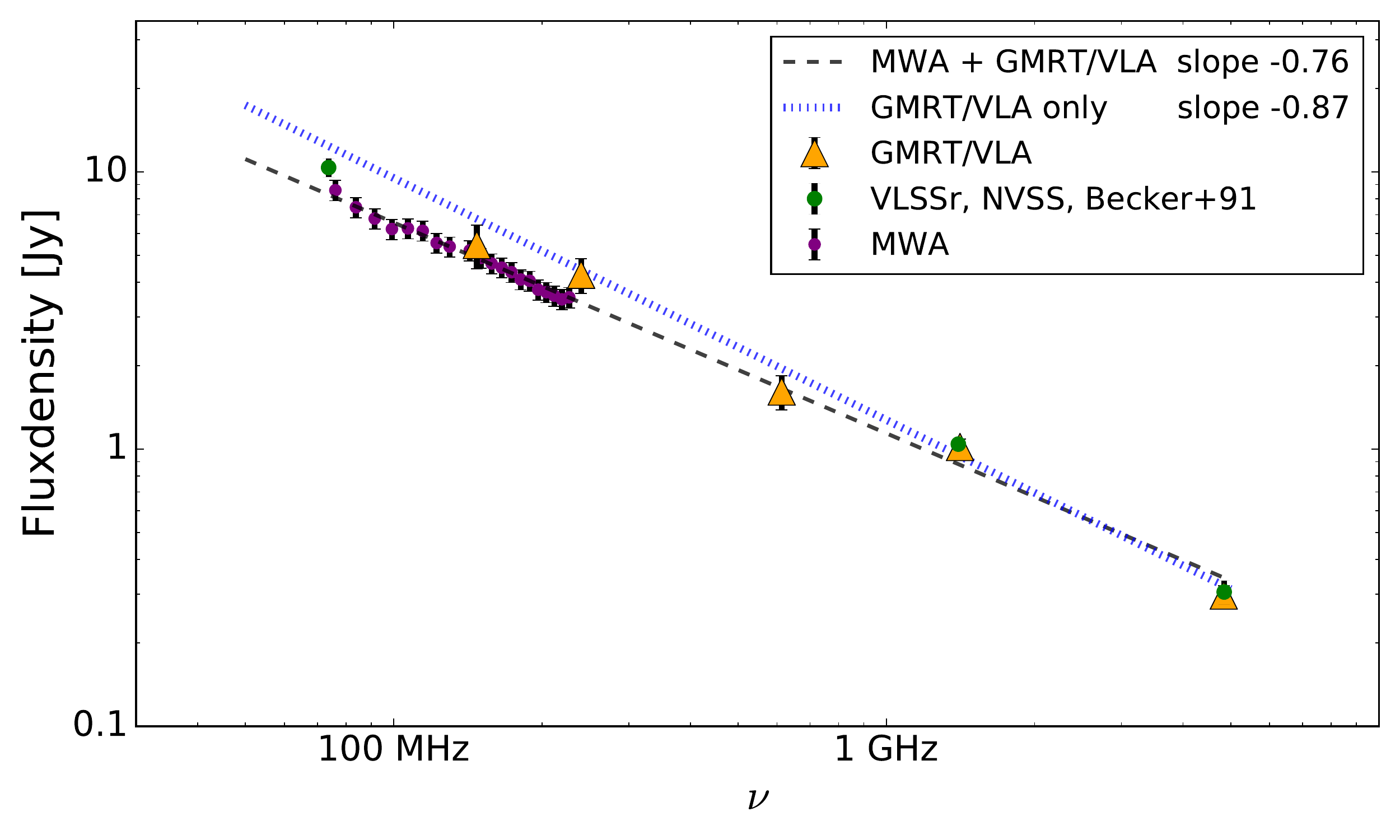}}
		\caption{Integrated radio spectrum of NGC\,741.}
		\label{fig:spectrum}
	\end{figure*}

	The flux densities are summarized in Table \ref{tab:rflux} along with the fluxes from GLEAM,
	VLA Low-Frequency Sky Survey Redux (VLSS-R) at $\SI{74}{MHz}$ (\citealp{2014MNRAS.440..327L}), NVSS survey at
	$\SI{1.4}{GHz}$ (\citealp{1998AJ....115.1693C}), and the $\SI{4.85}{GHz}$ flux reported by \cite{1991ApJS...75....1B}.
	We note that our GMRT flux at $\SI{150}{MHz}$ ($5.44\pm0.82$ Jy) is significantly higher than the flux of $\SI{2.58(26)}{Jy}$ reported
	in the catalog from the TGSS Alternative Data Release (\citealp{2016arXiv160304368I}). By integrating the flux directly on the
	TGSS-ADR mosaic image, we find a slightly higher value of $\SI{3.58(54)}{Jy}$, which, however, is still in disagreement
	with our flux at $\sim 1.9\,\sigma$ level and MWA flux ($\sim 2\,\sigma$) at $\SI{150}{MHz}$.
	The integrated radio spectrum is shown in Fig. \ref{fig:spectrum}.
	
	The data are given on the \cite{2012MNRAS.423L..30S} flux density scale, while the MWA data are given on the \cite{1977A&A....61...99B} flux density scale. As there are no calibrated flux density scales well suited for frequencies below 240\,MHz (\citealp{2016arXiv160905940P}), we expect only a small difference between these scales ($\leq 2\%$). We do not attempt to correct to a single flux scale. Nevertheless, we note that the GMRT spectrum shows broad agreement with a single power-law model with an overall slope of $\num{0.76(1)} $ between $\SI{74}{MHz}$ and $\SI{5}{GHz}$.
	
	\subsubsection{Spectral study of the south-west tail}
	\label{ch:swtail}
\begin{figure}
		\centering
		\resizebox{0.99\hsize}{!}{\includegraphics[clip,trim=50px 25px 0px 30px,width=0.98\linewidth]{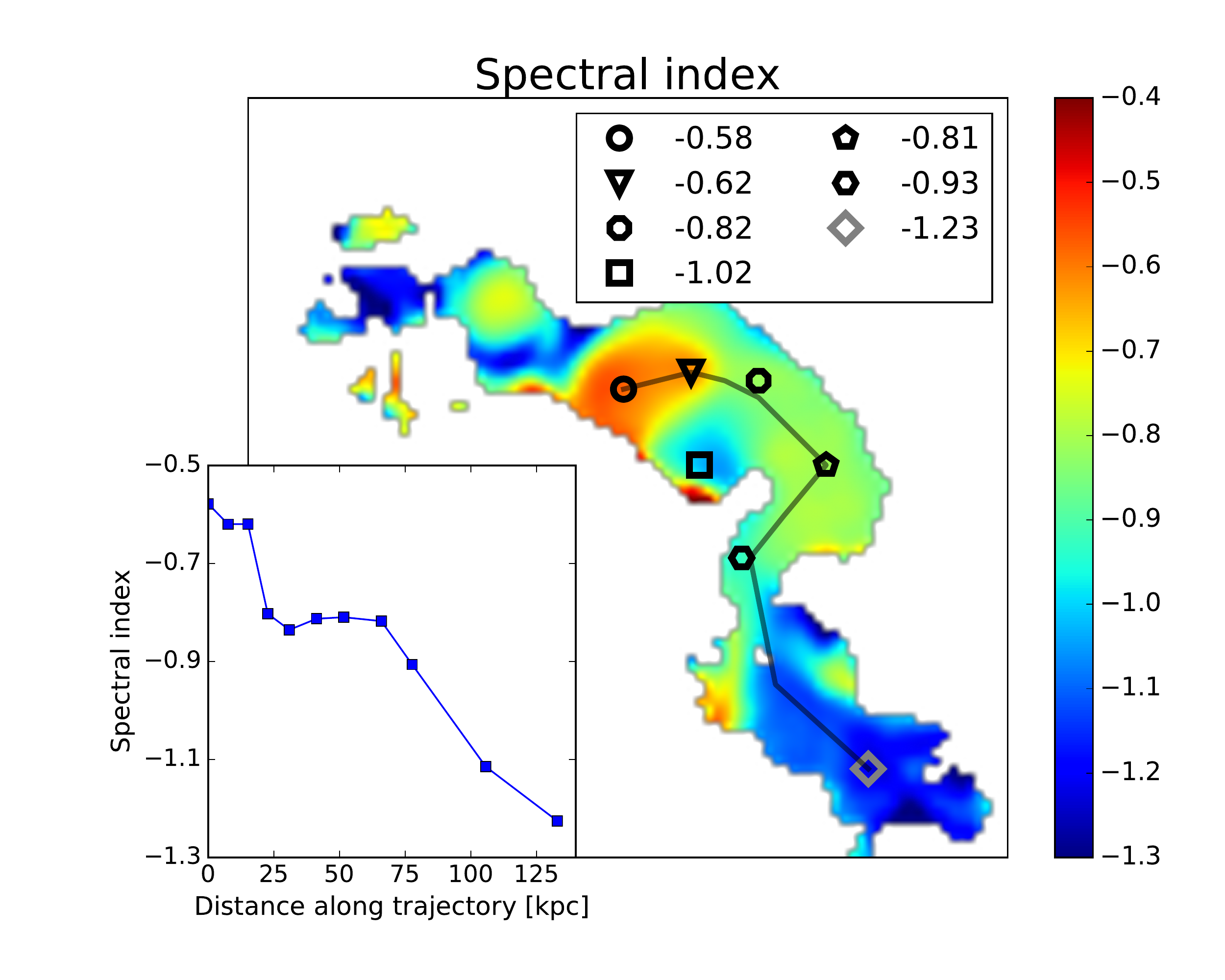}}
		\caption{Radio spectral index map using the GMRT and VLA data from 150\,MHz to 4.8\,GHz (see Section \ref{ch:flux_comp}). The markers show regions of interest and the legend lists the spectral index at that point. The lower left panel shows the spectral index along the assumed trajectory of NGC\,742.}
		\label{fig:spec_indx}
	\end{figure}

\begin{figure}
		\centering
		\resizebox{0.99\hsize}{!}{\includegraphics[clip,trim=140px 50px 0px 30px,width=0.98\linewidth]{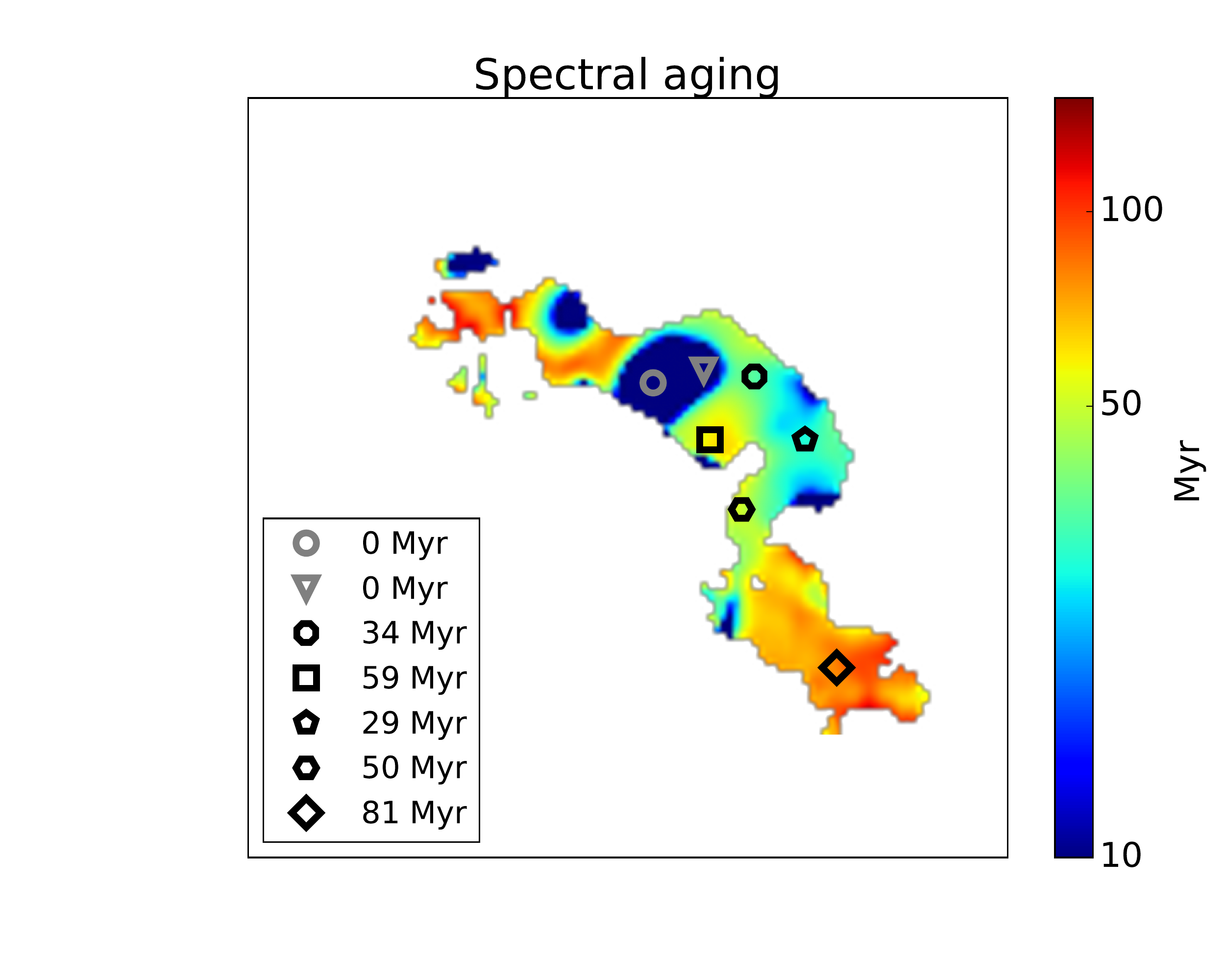}}
		\caption{Spectral aging map derived from the calculation of the break frequency (Eq. \ref{eq:aging}) assuming a power-law slope of $\num{-0.76}$.}
		\label{fig:aging}
	\end{figure}

	For a detailed analysis of the south-west radio tail using the GMRT and VLA imaging data (150\,MHz to 4.8\,GHz), we homogenized all images which have different resolutions and UV-coverage using the \verb|CASA| software.
	All images have been produced using the same uv-range, Gaussian-smoothed to $\SI{25}{arcsec}$ (the beam of the lowest resolution image), and re-gridded.
	In Fig. \ref{fig:spec_indx} and Fig. \ref{fig:aging} we show all regions where a spectral index could be computed (i.e., regions where the flux exceeds twice the noise level at all frequencies).
	The circle and triangle mark the positions of NGC\,742 and NGC\,741, respectively. The other points give the spectral index and synchrotron age of different interesting regions.
	The spectral age is computed from the break frequency obtained by fitting a power-law spectrum with an exponential cut-off as in the JP model (Jaffe-Perola, see e.g., \citealp{1991ApJ...383..554C}). Following \cite{2001AJ....122.1172S}, the age is then calculated from:
	
	\begin{equation}
	\label{eq:aging}
	\frac{t_\mathrm{age}}{\si{Gyr}} = \num{1.59} \frac{\sqrt{B} }{B^2 + B_\mathrm{CMB}^2} \left( \left(1+z \right)\frac{\nu_\mathrm{break}}{\SI{1}{GHz}} \right)^{-\frac{1}{2}}~,
	\end{equation}
	where the magnetic field $B$ is given in $\si{\mu G}$ and the inverse Compton losses are accounted for with the equivalent magnetic field of the CMB,
	$B_\mathrm{CMB} = \num{3.25} \left(1+z\right)^2 \si{\mu G}$. We assume $B = \SI{2}{\mu G}$, but the dependence of the age on the magnetic field strength is weak
	($\SI{1}{\mu G}$ decreases the age by 5\%, while $\SI{5}{\mu G}$ decreases the age by 28\% with respect to our default choice).
	To better constrain the spectral age we assume that the power-law component has a fixed slope of $\num{-0.76}$, which agrees with the radio spectrum of the integrated source.
	
	The flattest spectrum is located around NGC\,742, as also indicated by the high resolution images in the previous section. We also detect that along the south-west radio tail, the spectral index shows a clear steepening (up to an index of $\num{-1.2}$ to $\num{-1.3}$). 
	The general shape of the south-west tail cannot be easily explained, for example by gravitational bending or galactic motion, but was probably caused by gas sloshing in the ICM which perturbed the end of the tail (and also the older particle population) to the south. Similar shapes have also been seen in other head-tail radio galaxies (e.g., for 3C 465 see \citealp{1984ApJ...278...37E}) and the bent radio jet in NGC5044 by \citealp{2009ApJ...705..624D}.

	\subsection{Optical}
	
\begin{figure}
		\centering
		\resizebox{0.99\hsize}{!}{\includegraphics[clip,trim=0px 50px 0px 50px,width=0.9\linewidth]{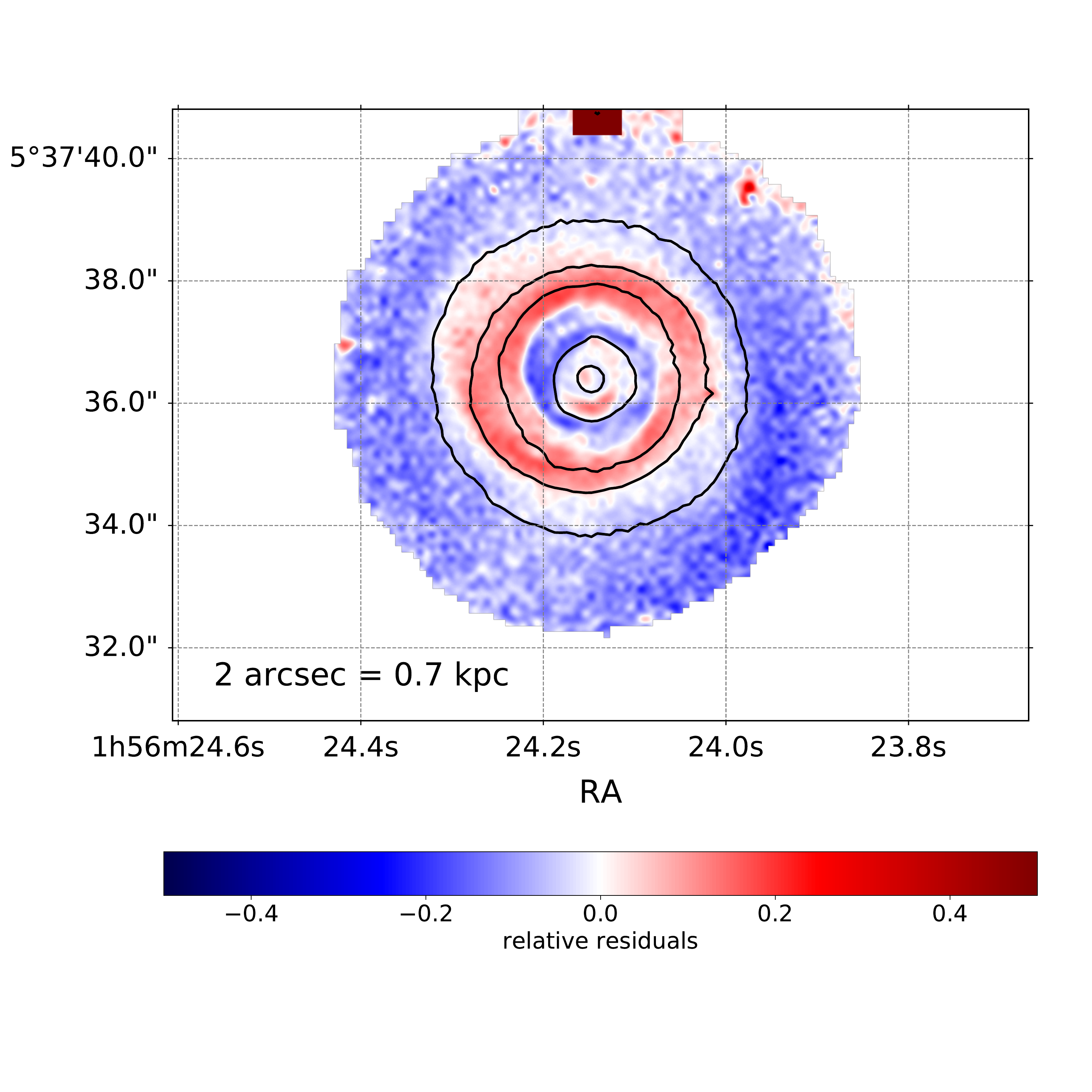}}
		\caption{NGC\,742 Residuals of the HST image $\SI{555}{nm}$ after subtracting the best-fit 2D Sersic model. Red (blue) regions have a higher (lower) flux than the model prediction. Contour lines follow the optical surface brightness.}
		\label{fig:hst}
	\end{figure}

	We use the available HST data at $\SI{555}{nm}$, taken with the Wide-Field Planetary Camera 2 (WFPC2) from the Hubble Legacy Archive (HST Proposal 6587), to fit 2D Sersic models to the two galaxies NGC\,741 and NGC\,742 with the addition of a constant background (using \verb|sherpa|). Both galaxies have a Sersic index of about 1 (1.2 and 1.6) as expected for giant elliptical galaxies.
	Subtracting these 2D Sersic models from the HST image shows there are no  significant surface brightness residuals near NGC\,741, but there are clear residuals in the form of rings around the center of NGC\,742 (see Fig. \ref{fig:hst}).
	The 2D Sersic model allows for ellipticity, but the eccentricity is only 1\% and consistent with 0. 
	The red ring in Fig. \ref{fig:hst} may be due to the recent encounter with NGC\,741 or the recent accretion of a gas rich dwarf galaxy. The formation of rings after a close encounter has been seen in literature, even for early type galaxies (\citealp{2010ApJS..186..427N,2012MNRAS.420.1158M,2017A&A...598A..45G}).
	\section{Discussion}\label{sec:disc}
	\subsection{Gas cooling}
	To investigate the importance of gas cooling and thermal instabilities in the group center, we computed the ratio of the cooling time to the free-fall time, $\frac{t_\mathrm{cool}}{t_\mathrm{ff}}$, as suggested, e.g., by \cite{2015Natur.519..203V} or \cite{2016arXiv161004617H}. At a distance of $\SI{18}{kpc}$ from the group center (approximately the distance to NGC\,742), the ratio is about $\num{100}$, while in the center of the group near $\SI{1}{kpc}$, the ratio drops to approximately 10 (see the values in Tab. \ref{tab:gascool}). This corresponds to the threshold were the growth of thermal instabilities due to cooling becomes significant.
	Several other basic ICM parameters are also summarized in Tab. \ref{tab:gascool}.
	We conclude for our group that the amount of cold gas within the region of the BCG and its supermassive black hole is not refilled fast enough to allow star formation (SF) to be initiated. There is no available $H\alpha$ data, but we can estimate the SF rate from the WISE data (using W3 and W4 bands as in \citealp{2013ApJ...774...62L,2013AJ....145....6J}). Compared to studies like, e.g., \cite{2016ApJ...817...86M}, we get a fairly low SF rate for NGC\,741 of about $\SI{0.25}{M_\odot\,yr^{-1}}$.
	
	\begin{deluxetable*}{CCCCCC}
		\tabletypesize{}
		\tablecaption{Gas properties at small radii round the galaxy NGC\,741. \label{tab:gascool}}
		\tablehead{
			\colhead{$R$}& \colhead{$t_\mathrm{cool}$}& \colhead{$t_\mathrm{cool} / t_\mathrm{ff}$} & \colhead{$n_\mathrm{e}$} & \colhead{$K$} & \colhead{$\dot M_\mathrm{classic}$}\\
			$[\mathrm{kpc}]$ &$[\si{Gyr}]$ &  &$[\si{cm^{-3}}]$ &$[\si{keV\,cm^2}]$ & $[\si{\frac{M_\odot}{yr}}]$
		}
		\colnumbers
		\startdata
		1 & \num{0.45(4)} & \num{12(5)} & \num{9.2(5)e-3} & \num{12(1)} & \num{2.4(3)e-3} \\
		10 & \num{3.05(16)} & \num{50(6)} & \num{2.5(1)e-3} & \num{45(2)} & \num{8.1(6)e-2} \\
		50 & \num{60(5)} & \num{483(36)} & \num{3.6(2)e-4} & \num{317(16)} & \num{1.6(2)e-1} \\
		\enddata
		\tablenotetext{}{(1) Distance from core of NGC\,741. (2) Cooling time at $R$. (3) Ratio of the cooling and free-fall time indicating thermal instabilities. (4) Electron number density from model. (5) Entropy. (6) Classical mass deposition rate.}
		
	\end{deluxetable*}

	The entropy profile derived from the deprojected temperature profile has a relatively steep slope of  $\num{1.35(2)}$, an entropy of $\SI{796(5)}{keV\,cm^2}$ at $\SI{100}{kpc}$, and an extrapolated central entropy of $\SI{9(2)}{keV\,cm^2}$. In general, galaxy groups show a larger scatter in the slope of the entropy profile compared to rich clusters (\citealp{2003ApJ...598..250S,2009ApJS..182...12C}).
	Since the slope of the entropy profile is steeper than that expected from purely gravitational heating (slope of 1.1; \citealp{2001ApJ...546...63T}), the outer regions of the group must have been strongly affected by the central AGN. Examining the gas mass fraction as a function of radius, we find a slight decrease in the region between $20$ and $\SI{100}{kpc}$. Together with the steep entropy profile, we conclude that, at intermediate scales, the galaxy group is relatively gas poor and cannot cool efficiently. In addition, the cooling time near $\SI{50}{kpc}$ is about five times the Hubble time, much more than expected for a group (e.g., \citealp{2006MNRAS.372.1496S}). This is the likely reason for the temperature peak between $1$ and $\SI{2}{arcmin}$.
	
	\begin{figure*}
		\centering
		\resizebox{0.99\hsize}{!}{\includegraphics[clip,trim=50px 50px 0px 30px,width=0.9\linewidth]{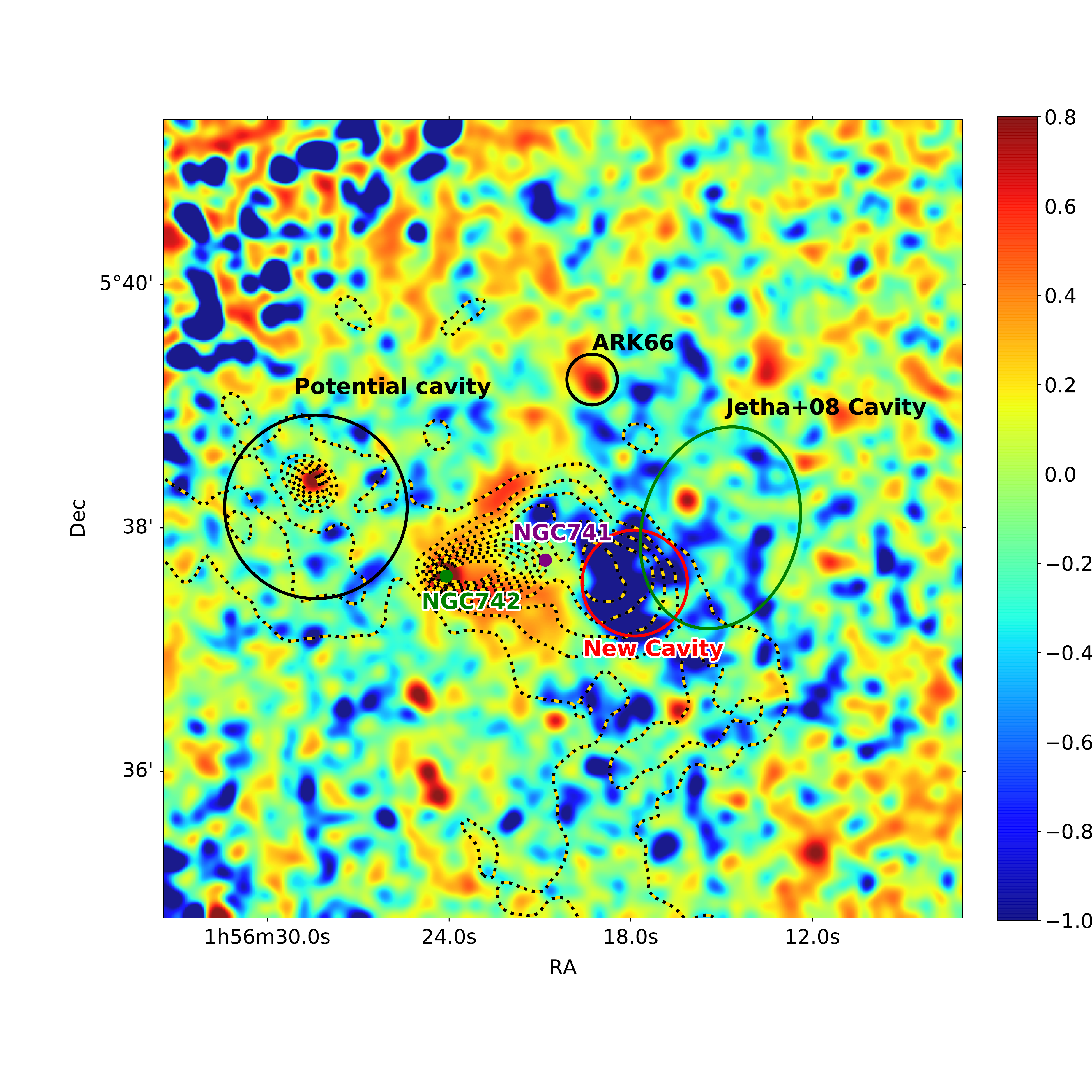}}
		\caption{Residual image with respect to the double beta 2D model fit. The color scale shows the fractional residuals where positive values refer to an underestimated flux in the model. The dotted lines are the $3\sigma$ radio contours from $\SI{235}{MHz}$ GMRT observations. $\SI{1}{arcmin} = \SI{21}{kpc}$}
		\label{fig:rel_resid}
	\end{figure*}
	
	\subsection{Cavities}
	\label{ch:cavity}
	
	\subsubsection{Jetha+08 cavity}
	
	In our deep X-ray observations we do not clearly detect the previously claimed X-ray cavity to the west of NGC\,741 (\citealp{2008MNRAS.384.1344J}), marked in Fig. \ref{fig:rel_resid}. We estimate the decrease in number of counts assuming the cavity region (modeled as a sphere for simplicity) is completely depleted of thermal gas. Assuming our double-beta parametrization of the surface brightness profile, we estimate 731 counts from the ICM along the line of sight to the cavity, of which 105 counts originate from within the cavity volume. Including about 530 background counts in the calculation (total $\SI{1261(36)}{cts}$), a completely empty cavity would be significant at a $3\sigma$ level. The fact that we measure exactly $\SI{1254}{cts}$ within the cavity indicates that there is no surface brightness decrease, assuming our model is correct.

	Nevertheless, we calculate the enthalpy of the claimed cavity from the density and temperature profiles using
	\begin{equation}
	H = \frac{\gamma}{\gamma -1 } pV~,
	\end{equation}
	where $\gamma = \frac{4}{3}$ for a fully relativistic gas, and $p$ is the pressure within the volume $V$.
	We find a value of $\SI{2.1(2)e58}{erg}$.
	In order to determine the nature of the any existing cavity, we compare, as in \cite{2014MNRAS.444.1236P}, the cooling luminosity, inferred within a region where the cooling time is shorter than $\SI{3}{Gyr}$ (see \citealp{2014MNRAS.438.2341P}), in our case $\SI{10}{kpc}$, with the enthalpy divided by the sound crossing time $t_\mathrm{s}$.
	The spectral analysis gives us a cooling luminosity of $\SI{2.59(4)e41}{erg\,s^{-1}}$.
	With the sonic timescale, $t_\mathrm{s} = \SI{27}{Myr}$, we estimate cavity power to be $\SI{2.6e43}{erg\,s^{-1}}$, which is two orders of magnitude above the measured luminosity.
	We conclude that the claimed cavity is too large and too close to the core of the BCG. It cannot be a young structure still being powered by the AGN jets. We therefore cannot confirm the existence of the ghost cavity, and it seems likely that it was misidentified owing to the asymmetry of the surface brightness distribution.

	\subsubsection{Small west cavity}
	\label{ch:smallcavity}
	We find in the normalized residual image (Fig. \ref{fig:rel_resid}) a strong decrease in surface brightness ($25\sigma$ with respect to the surrounding annulus) about $\SI{16}{kpc}$ west of NGC\,741. This seems to be a very good candidate for a cavity inflated by the NGC\,741 AGN, especially because the radio emission seems to roughly overlay the feature seen in the X-rays. Within the circular region ($\SI{8}{kpc}$ radius) of the cavity we measure
	$\SI{397}{cts}$, while we would expect $\SI{702}{cts}$ from the surface brightness and density profiles.
	A completely empty sphere at this position would reduce the number of counts only by $\SI{69}{cts}$, not enough to explain the observed deficit. If we assume an ellipsoidal cavity elongated along the line of sight, a semi-major axis of $\SI{37}{kpc}$ is required, $4.6$ times the size of the other axes. 
	
	The enthalpy and cavity power (based on the sound crossing timescale) of this newly detected spherical under-dense region is $\SI{4.9(5)e57}{erg}$ and $\SI{5.1(5)e42}{erg\,s^{-1}}$, which would require $\SI{600}{Myr}$ to be consistent with the $4pV$ relation (\citealp{2014MNRAS.444.1236P}).

	\subsubsection{Potential east cavity}
	The interaction between NGC\,742 and NGC\,741 (see Section \ref{ch:ngc742merging}) might have affected the position of NGC\,741. We take the central velocity dispersion measurements for both galaxies from \cite{2012ApJ...756..117F} and use the relations from \cite{2011Natur.480..215M} and \cite{2016arXiv160704275Z} to get
	the stellar masses ($\SI{3.6e11}{M_\odot}$ for NGC\,741, $\SI{5e10}{M_\odot}$ for NGC\,742). Using this mass ratio of about $7$ to $1$, we conclude that a shift of NGC\,741 of about $10$ to $\SI{20}{arcsec}$ to the west (i.e., it's current position) is plausible. Given the presence of the newly identified small western cavity, we might expect a second cavity on the eastern side of NGC\,741, located further from the galaxy. The two cavities would have been roughly equidistant from NGC\,741 before the encounter with NGC\,742.
	
	We do indeed see extended radio emission to the east of NGC\,741, especially at $\SI{235}{MHz}$ and $\SI{610}{MHz}$ ("Potential cavity" in Fig. \ref{fig:rel_resid}).
	This emission could be due to an old radio lobe inflated by NGC\,741, although there is no emission at the same radius on the west side of NGC\,741. Constructing a spectral index map from those two frequencies only (because of the high resolution), we find overall a rather steep index ($>2$), with a few regions close to the flatter spectrum of NGC\,742. It seems that part of the emission in this region might be powered by the AGN of NGC\,742, with the remainder arising from an older outburst of NGC\,741.
	We do not see indications at this position for an X-ray surface brightness decrease. This might have been obscured by the shock produced by NGC\,742.
	
	\begin{figure}
		\centering
		\resizebox{0.99\hsize}{!}{\includegraphics[width=0.9\linewidth]{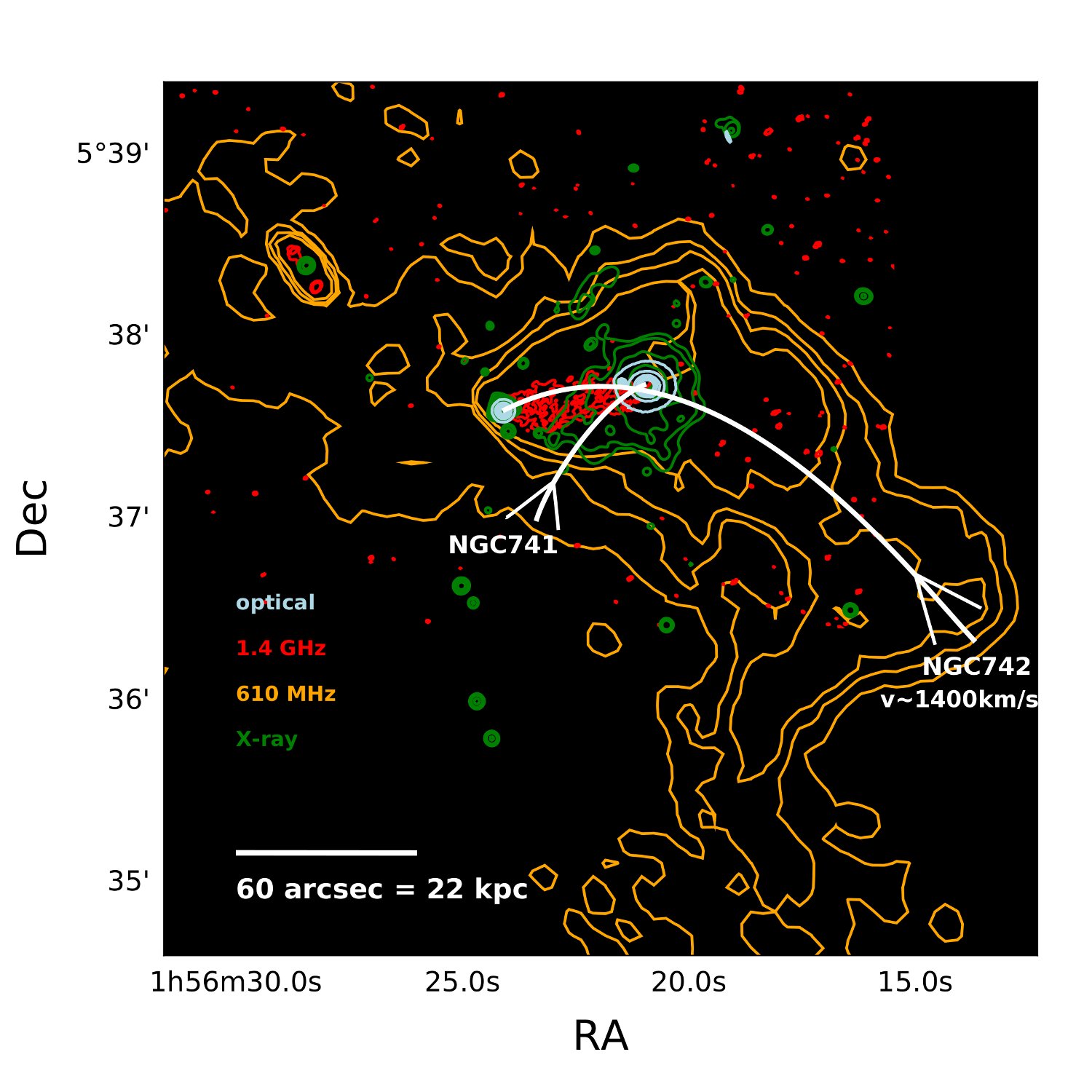}}
		\caption{Illustration of the merging scenario of NGC\,742 with the galaxy group.}
		\label{fig:scenario}
	\end{figure}
	
	\subsection{NGC\,742 merging event}
	\label{ch:ngc742merging}
	Our suggested scenario for the infall of NGC\,742 into the NGC\,741 group is summarized in Fig. \ref{fig:scenario}. We conclude from the spectral index map of the south-west radio tail that NGC\,742 approached the galaxy group with an approximate velocity of $\SI{1400}{km\,s^{-1}}$. Its path is characterized by the remnant radio emission of the AGN, although the shape some of the more distant regions of the south-west tail might have been disturbed by gas motions in the surrounding ICM.
	The surface brightness deficiency (classified as a smaller cavity in Section \ref{ch:smallcavity}) might also be caused by the encounter of NGC\,742 dragging gas away from this region. The fact that the deficit in counts in this region is larger than expected for an AGN inflated bubble, might point out that it is a superposition of multiple effects.
	
	The close encounter with the BCG and the much denser ICM of NGC\,741 caused the stripping of any gas associated with NGC\,742 and may have also displaced some of the gas from the group core. The activity of the NGC\,742 AGN may also have have been enhanced, leading to the prominent head-tail structure.
	If one assumes that the shape of this head-tail source resembles a Mach cone, the velocity can be estimated from the cone-angle to be $\sim\SI{1300}{km\,s^{-1}}$, in good agreement with the velocity estimated from the spectral index map.
	The velocity of a head-tail radio source in a galaxy cluster can also be estimated from the radio jet curvature  and cylindrical jet diameter (see e.g., \citealp{2011ApJ...738..145F} or simulations by \citealp{2013MNRAS.431..781M}). We assume a jet velocity on the order of $0.1 c$, and a ratio of jet curvature and cylindrical diameter of $3.75$, and find a galaxy velocity of order $\SI{500}{km\,s^{-1}}$, less than our other estimates. 
	If we look at the optical recession velocities of NGC\,741 and NGC\,742 (\citealp{2004ApJ...607..202M}) we find a difference of $\SI{403}{km\,s^{-1}}$. We assume this is the line-of-sight velocity component of NGC\,742 relative to the galaxy group, while the projected velocity component is estimated from the trajectory ($\SI{130}{kpc}$) and the spectral aging $\SI{90}{Myr}$ to be $\SI{1410}{km\,s^{-1}}$. This leads to a 3D velocity of $\SI{1466}{km\,s^{-1}}$. This is in good agreement with the infall velocity due to the gravitational potential of the galaxy group if the infall of NGC\,742 started at around $R_{500}$ until the current position, $\SI{18}{kpc}$ away from the group center.

	In the X-ray regime we reported several unexpected features for a cool core cluster or group:
	Perhaps the most notable is the X-ray bright filament between the core of NGC\,741 and NGC\,742, which consists of low entropy gas and is consistent with being stripped from NGC\,742.
	Assuming a cylindrical shape, our spectral analysis of the X-ray data using an MCMC reveals that the filament is in rough pressure equilibrium with the surrounding medium ($\SI{4.3(9)e-11}{dyn\,cm^{-2}}$ in the filament and $\SI{3.4(2)e-11}{dyn\,cm^{-2}}$ in the surrounding).
	The filament could also be gas from the core of NGC\,741, drawn outwards during the close encounter. The X-ray emission from the filament is slightly offset from the bent radio lobes of NGC742 (see Fig. \ref{fig:scenario}). The origin of the radio emission is attributed to the strongly bent jets of the AGN of NGC\,742 and so it is plausible that it is not perfectly aligned with the X-ray emission. More evidence for a recent disruptive event in NGC\,742 is provided by the rings visible in the optical image (Fig. \ref{fig:hst}).
	
	In Figure \ref{fig:rel_resid} we see a V-shaped structure in the X-ray surface brightness residuals to the east of NGC\,742. The position and shape of this feature might suggest that there is a cold or shock front present.
	We also see a temperature discontinuity in this region in Fig. \ref{fig:maps}. Taking spectra from the region of the increased surface brightness, and from the region to its east, shows that in the east, the temperature is about 75\% higher, as would be expected for a cold front with a Mach number of about $1.7$.
	Using the sound speed at this location, this Mach number implies a velocity for NGC\,742 of $\SI{1100(66)}{km\,s^{-1}}$.
	This estimate is about 30\% lower than the velocity from spectral aging, but given the many uncertainties in selecting the (pre-)shock region for the spectral extraction, and the assumptions in the radio spectral aging calculation, it still clearly confirms a supersonic motion.

	\subsection{Other merging activity}
	To the south-west of NGC\,741 ($\sim \SI{30}{arcsec}$ distance) we see a two arm-like features in the soft X-ray image. The first one, around $\SI{18}{kpc}$ length, is well aligned to the edge of the south-west radio-tail, while the other, approximately $\SI{16}{kpc}$ in length, is perpendicular to the first arm, pointing to the south. The origin of these features remains unclear, although they are likely related to the perturbation caused by the infalling galaxy NGC\,742 (e.g., gas pushed aside during the encounter).
	In Fig. \ref{fig:xmm} (left) one can see that this X-ray filament is aligned with the radio emission tracing the infall of NGC\,742.
	
	We detect X-ray emission from other group member galaxies, including one with an X-ray tail. Although we do not see other head-tail radio sources, this clearly indicates further merging activity in this system. $\SI{35}{arcsec}$, or $\SI{13}{kpc}$, to the north-east of the core of NGC\,741 we see a filament in the soft X-ray band. Extracting a spectrum of this region reveals that its temperature is consistent with the surrounding ICM. Although the abundance might be slightly higher (30\%) in the region of the north-east filament, with the number of available counts we cannot be sure that this difference is statistically significant. Nevertheless, this region could be the remnant of an earlier merging interaction, i.e. ram pressure stripped gas. The sound-crossing time of the north-east filament is about $\SI{13}{Myr}$.
	
	\section{Summary and Conclusions}\label{sec:conclusion}
	We analyzed the kinematic and thermodynamic structure of the galaxy group NGC\,741. This nearby group is caught in an unusual dynamical state and exhibits a recent merging event, with a member galaxy passing through the core region close to the BCG. Our results exploit data from recent long Chandra and XMM-Newton observations as well as from GMRT and VLA covering five frequencies between $\SI{150}{MHz}$ and $\SI{5}{GHz}$. We also include constraints from the MWA GLEAM survey and optical HST observations in drawing our conclusions, which can be summarized as follows:
	\begin{itemize}
		\item Our X-ray analysis shows that the group has a cool-core with a minimum temperature of about $\SI{0.5}{keV}$. The peak temperature of the ICM, about $\SI{2}{keV}$, is reached at $\SI{1.5}{arcmin}$ radius, and the total hydrostatic mass $M_{500} = \SI{5.2(4)e13}{M_\odot}$. The gas mass fraction is only about 3\% at $R_{2500}$.
		\item We find several merging or infalling objects. Most prominent is NGC\,742, only about $\SI{18}{kpc}$ in projection from the BCG. We detect three other X-ray bright galaxies, one of them with a tail indicative of gas stripping caused by its infall through the group. Residual gas filaments to the north-east of the BCG may indicate other past merger events.
		\item By developing a new algorithm based on the Iron-L line position (see also \citealp{2009ApJ...705..624D}) and the relative line flux, we are able to construct temperature and relative abundance maps even with limited numbers of counts. These maps show a hotter temperature to the east and cooler structure to the south of NGC\,741. Along with the higher abundance in the center of the group, we see several peaks in the abundance map (confirmed using standard spectroscopic analysis), probably due to the stripped gas of merging galaxies, especially along the inferred path of NGC\,742.
		\item The X-ray filament connecting NGC\,742 and NGC\,741 may itself be an X-ray tail, since its entropy is consistent with being gas stripped from NGC\,742. High-resolution radio images also show a head-tail structure, slightly offset from the X-ray filament. We attribute the radio emission to the strongly bent jets of the AGN in NGC\,742. The optical HST image reveals irregularities in the stellar component of the galaxy, which could be witnesses of the past interaction with the BCG.
		\item The large-scale radio emission shows a long (\mbox{$>\SI{130}{kpc}$}) tail to the south-west of the BCG, which we interpret as older emission from the AGN of NGC\,742. The radio spectral index map shows a clear steepening from a power-law index of $-0.6$ around NGC\,742, to $-1.3$ and beyond at the end of the tail. We map the estimated spectral age along this tail and conclude that the time of last injection of energetic particles at the end of the tail was around $\SI{90}{Myr}$ ago.
		\item The ghost cavity to the west of the BCG cannot be confirmed. We discover a cavity at a closer distance, about $\SI{16}{kpc}$ from NGC\,741. This region shows a much stronger surface brightness decrease and exhibits non-thermal radio emission. The depth of the surface brightness deficit suggests that this cavity is very elongated (factor of $4.5$) along the line of sight. 
	\end{itemize}
	
	Given the large amount of multi-wavelength data available for this object, we are able to construct a detailed scenario for the history of the NGC\,741 -- NGC\,742 interaction.
	Some uncertainties remain, particularly regarding the shape of the south-west radio tail, and the unexpected irregularity of the spectral steepening along its length. We conclude that the galaxy had a strong impact on the chemical composition of the galaxy group, and that since the interaction happened fairly recently, the creation of prominent gas-sloshing features can be expected in the future as the group core responds to the impetus imparted by the interaction.
	
	\section*{Acknowledgements}
	The authors acknowledge the comments provided by the anonymous referee, which helped to improve this paper.
	The authors would like to thank Mark Birkinshaw, Diana Worrall, Reinout van Weeren and Tracy Clarke for helpful discussions.
	Support for this work was
	provided by the National Aeronautics and Space Administration
	(NASA) through Chandra Award Number GO3--14143X and GO5--16135X issued
	by the Chandra X-ray Observatory Center (CXC), which is
	operated by the Smithsonian Astrophysical Observatory (SAO)
	for and on behalf of NASA under contract NAS8--03060.
	Basic research in radio astronomy at the Naval Research Laboratory is supported by 6.1 base funding.
	MJ-H and SD are funded by a Marsden Fund Grant (PI Johnston-Hollitt) administered by the Royal Society of New Zealand.
	This research made use of Astropy, a community-developed core Python package for Astronomy.
	
	\bibliographystyle{aasjournal}
	\bibliography{Astro}

\begin{thebibliography}{}
\expandafter\ifx\csname natexlab\endcsname\relax\def\natexlab#1{#1}\fi

\bibitem[{Allen(1995)}]{1995MNRAS.276..947A}
Allen, S.~W. 1995, \mnras, 276, 947

\bibitem[{{Asplund} {et~al.}(2009){Asplund}, {Grevesse}, {Sauval}, \&
  {Scott}}]{2009ARA&A..47..481A}
{Asplund}, M., {Grevesse}, N., {Sauval}, A.~J., \& {Scott}, P. 2009, \araa, 47,
  481

\bibitem[{Baars {et~al.}(1977)Baars, Genzel, Pauliny-Toth, \&
  Witzel}]{1977A&A....61...99B}
Baars, J. W.~M., Genzel, R., Pauliny-Toth, I. I.~K., \& Witzel, A. 1977, \aap,
  61, 99

\bibitem[{Becker {et~al.}(1991)Becker, White, \& Edwards}]{1991ApJS...75....1B}
Becker, R.~H., White, R.~L., \& Edwards, A.~L. 1991, \apjs, 75, 1

\bibitem[{{Bharadwaj} {et~al.}(2014){Bharadwaj}, {Reiprich}, {Schellenberger},
  {Eckmiller}, {Mittal}, \& {Israel}}]{Bharadwaj2014}
{Bharadwaj}, V., {Reiprich}, T.~H., {Schellenberger}, G., {et~al.} 2014, \aap,
  572, A46

\bibitem[{{Bhattacharya} {et~al.}(2013){Bhattacharya}, {Habib}, {Heitmann}, \&
  {Vikhlinin}}]{2013ApJ...766...32B}
{Bhattacharya}, S., {Habib}, S., {Heitmann}, K., \& {Vikhlinin}, A. 2013, \apj,
  766, 32

\bibitem[{{Birkinshaw} \& {Davies}(1985)}]{1985ApJ...291...32B}
{Birkinshaw}, M., \& {Davies}, R.~L. 1985, \apj, 291, 32

\bibitem[{{B{\^i}rzan} {et~al.}(2004){B{\^i}rzan}, {Rafferty}, {McNamara},
  {Wise}, \& {Nulsen}}]{2004ApJ...607..800B}
{B{\^i}rzan}, L., {Rafferty}, D.~A., {McNamara}, B.~R., {Wise}, M.~W., \&
  {Nulsen}, P.~E.~J. 2004, \apj, 607, 800

\bibitem[{Briggs(1995)}]{briggsPhD}
Briggs, D.~S. 1995, PhD thesis, New Mexico Institute of Mining Technology,
  Socorro, New Mexico, USA

\bibitem[{{Buote}(2000)}]{2000MNRAS.311..176B}
{Buote}, D.~A. 2000, \mnras, 311, 176

\bibitem[{Carilli {et~al.}(1991)Carilli, Perley, Dreher, \&
  Leahy}]{1991ApJ...383..554C}
Carilli, C.~L., Perley, R.~A., Dreher, J.~W., \& Leahy, J.~P. 1991, \apj, 383,
  554

\bibitem[{{Cash}(1979)}]{1979ApJ...228..939C}
{Cash}, W. 1979, \apj, 228, 939

\bibitem[{{Cavagnolo} {et~al.}(2009){Cavagnolo}, {Donahue}, {Voit}, \&
  {Sun}}]{2009ApJS..182...12C}
{Cavagnolo}, K.~W., {Donahue}, M., {Voit}, G.~M., \& {Sun}, M. 2009, \apjs,
  182, 12

\bibitem[{{Condon} {et~al.}(1998){Condon}, {Cotton}, {Greisen}, {Yin},
  {Perley}, {Taylor}, \& {Broderick}}]{1998AJ....115.1693C}
{Condon}, J.~J., {Cotton}, W.~D., {Greisen}, E.~W., {et~al.} 1998, \aj, 115,
  1693

\bibitem[{{David} {et~al.}(2009){David}, {Jones}, {Forman}, {Nulsen},
  {Vrtilek}, {O'Sullivan}, {Giacintucci}, \&
  {Raychaudhury}}]{2009ApJ...705..624D}
{David}, L.~P., {Jones}, C., {Forman}, W., {et~al.} 2009, \apj, 705, 624

\bibitem[{{Eckmiller} {et~al.}(2011){Eckmiller}, {Hudson}, \&
  {Reiprich}}]{2011A&A...535A.105E}
{Eckmiller}, H.~J., {Hudson}, D.~S., \& {Reiprich}, T.~H. 2011, \aap, 535, A105

\bibitem[{{Eilek} {et~al.}(1984){Eilek}, {Burns}, {O'Dea}, \&
  {Owen}}]{1984ApJ...278...37E}
{Eilek}, J.~A., {Burns}, J.~O., {O'Dea}, C.~P., \& {Owen}, F.~N. 1984, \apj,
  278, 37

\bibitem[{{Eke} {et~al.}(1998){Eke}, {Navarro}, \&
  {Frenk}}]{1998ApJ...503..569E}
{Eke}, V.~R., {Navarro}, J.~F., \& {Frenk}, C.~S. 1998, \apj, 503, 569

\bibitem[{{Focardi} \& {Malavasi}(2012)}]{2012ApJ...756..117F}
{Focardi}, P., \& {Malavasi}, N. 2012, \apj, 756, 117

\bibitem[{{Foster} {et~al.}(2012){Foster}, {Ji}, {Smith}, \&
  {Brickhouse}}]{2012ApJ...756..128F}
{Foster}, A.~R., {Ji}, L., {Smith}, R.~K., \& {Brickhouse}, N.~S. 2012, \apj,
  756, 128

\bibitem[{{Freeland} \& {Wilcots}(2011)}]{2011ApJ...738..145F}
{Freeland}, E., \& {Wilcots}, E. 2011, \apj, 738, 145

\bibitem[{George(2017)}]{2017A&A...598A..45G}
George, K. 2017, \aap, 598, A45

\bibitem[{{Giacintucci} {et~al.}(2011){Giacintucci}, {O'Sullivan}, {Vrtilek},
  {David}, {Raychaudhury}, {Venturi}, {Athreya}, {Clarke}, {Murgia},
  {Mazzotta}, {Gitti}, {Ponman}, {Ishwara-Chandra}, {Jones}, \&
  {Forman}}]{2011ApJ...732...95G}
{Giacintucci}, S., {O'Sullivan}, E., {Vrtilek}, J., {et~al.} 2011, \apj, 732,
  95

\bibitem[{{Hogan} {et~al.}(2017){Hogan}, {McNamara}, {Pulido}, {Nulsen},
  {Russell}, {Vantyghem}, {Edge}, \& {Main}}]{2016arXiv161004617H}
{Hogan}, M.~T., {McNamara}, B.~R., {Pulido}, F., {et~al.} 2017, \apj, 837, 51

\bibitem[{{Huchra} {et~al.}(1999){Huchra}, {Vogeley}, \&
  {Geller}}]{1999ApJS..121..287H}
{Huchra}, J.~P., {Vogeley}, M.~S., \& {Geller}, M.~J. 1999, \apjs, 121, 287

\bibitem[{Intema {et~al.}(2017)Intema, Jagannathan, Mooley, \&
  Frail}]{2016arXiv160304368I}
Intema, H.~T., Jagannathan, P., Mooley, K.~P., \& Frail, D.~A. 2017, \aap, 598,
  A78

\bibitem[{{Jarrett} {et~al.}(2013){Jarrett}, {Masci}, {Tsai}, {Petty},
  {Cluver}, {Assef}, {Benford}, {Blain}, {Bridge}, {Donoso}, {Eisenhardt},
  {Koribalski}, {Lake}, {Neill}, {Seibert}, {Sheth}, {Stanford}, \&
  {Wright}}]{2013AJ....145....6J}
{Jarrett}, T.~H., {Masci}, F., {Tsai}, C.~W., {et~al.} 2013, \aj, 145, 6

\bibitem[{{Jeltema} {et~al.}(2008){Jeltema}, {Binder}, \&
  {Mulchaey}}]{2008ApJ...679.1162J}
{Jeltema}, T.~E., {Binder}, B., \& {Mulchaey}, J.~S. 2008, \apj, 679, 1162

\bibitem[{{Jetha} {et~al.}(2008){Jetha}, {Hardcastle}, {Babul}, {O'Sullivan},
  {Ponman}, {Raychaudhury}, \& {Vrtilek}}]{2008MNRAS.384.1344J}
{Jetha}, N.~N., {Hardcastle}, M.~J., {Babul}, A., {et~al.} 2008, \mnras, 384,
  1344

\bibitem[{Konstantopoulos {et~al.}(2010)Konstantopoulos, Gallagher, Fedotov,
  Durrell, Heiderman, Elmegreen, Charlton, Hibbard, Tzanavaris, Chandar,
  Johnson, Maybhate, Zabludoff, Gronwall, Szathmary, Hornschemeier, English,
  Whitmore, Mendes~de Oliveira, \& Mulchaey}]{2010ApJ...723..197K}
Konstantopoulos, I.~S., Gallagher, S.~C., Fedotov, K., {et~al.} 2010, \apj,
  723, 197

\bibitem[{Lane {et~al.}(2014)Lane, Cotton, van Velzen, Clarke, Kassim,
  Helmboldt, Lazio, \& Cohen}]{2014MNRAS.440..327L}
Lane, W.~M., Cotton, W.~D., van Velzen, S., {et~al.} 2014, \mnras, 440, 327

\bibitem[{{Lavaux} \& {Hudson}(2011)}]{2011MNRAS.416.2840L}
{Lavaux}, G., \& {Hudson}, M.~J. 2011, \mnras, 416, 2840

\bibitem[{{Lee} {et~al.}(2013){Lee}, {Hwang}, \& {Ko}}]{2013ApJ...774...62L}
{Lee}, J.~C., {Hwang}, H.~S., \& {Ko}, J. 2013, \apj, 774, 62

\bibitem[{{Lovisari} {et~al.}(2015){Lovisari}, {Reiprich}, \&
  {Schellenberger}}]{Lovisari2015}
{Lovisari}, L., {Reiprich}, T.~H., \& {Schellenberger}, G. 2015, \aap, 573,
  A118

\bibitem[{{Mahdavi} \& {Geller}(2004)}]{2004ApJ...607..202M}
{Mahdavi}, A., \& {Geller}, M.~J. 2004, \apj, 607, 202

\bibitem[{Mapelli \& Mayer(2012)}]{2012MNRAS.420.1158M}
Mapelli, M., \& Mayer, L. 2012, \mnras, 420, 1158

\bibitem[{{McConnell} {et~al.}(2011){McConnell}, {Ma}, {Gebhardt}, {Wright},
  {Murphy}, {Lauer}, {Graham}, \& {Richstone}}]{2011Natur.480..215M}
{McConnell}, N.~J., {Ma}, C.-P., {Gebhardt}, K., {et~al.} 2011, \nat, 480, 215

\bibitem[{{McDonald} {et~al.}(2016){McDonald}, {Stalder}, {Bayliss}, {Allen},
  {Applegate}, {Ashby}, {Bautz}, {Benson}, {Bleem}, {Brodwin}, {Carlstrom},
  {Chiu}, {Desai}, {Gonzalez}, {Hlavacek-Larrondo}, {Holzapfel}, {Marrone},
  {Miller}, {Reichardt}, {Saliwanchik}, {Saro}, {Schrabback}, {Stanford},
  {Stark}, {Vieira}, \& {Zenteno}}]{2016ApJ...817...86M}
{McDonald}, M., {Stalder}, B., {Bayliss}, M., {et~al.} 2016, \apj, 817, 86

\bibitem[{{Mernier} {et~al.}(2015){Mernier}, {de Plaa}, {Lovisari}, {Pinto},
  {Zhang}, {Kaastra}, {Werner}, \& {Simionescu}}]{2015A&A...575A..37M}
{Mernier}, F., {de Plaa}, J., {Lovisari}, L., {et~al.} 2015, \aap, 575, A37

\bibitem[{{Morsony} {et~al.}(2013){Morsony}, {Miller}, {Heinz}, {Freeland},
  {Wilcots}, {Br{\"u}ggen}, \& {Ruszkowski}}]{2013MNRAS.431..781M}
{Morsony}, B.~J., {Miller}, J.~J., {Heinz}, S., {et~al.} 2013, \mnras, 431, 781

\bibitem[{Nair \& Abraham(2010)}]{2010ApJS..186..427N}
Nair, P.~B., \& Abraham, R.~G. 2010, \apjs, 186, 427

\bibitem[{{Panagoulia} {et~al.}(2014{\natexlab{a}}){Panagoulia}, {Fabian}, \&
  {Sanders}}]{2014MNRAS.438.2341P}
{Panagoulia}, E.~K., {Fabian}, A.~C., \& {Sanders}, J.~S. 2014{\natexlab{a}},
  \mnras, 438, 2341

\bibitem[{{Panagoulia} {et~al.}(2014{\natexlab{b}}){Panagoulia}, {Fabian},
  {Sanders}, \& {Hlavacek-Larrondo}}]{2014MNRAS.444.1236P}
{Panagoulia}, E.~K., {Fabian}, A.~C., {Sanders}, J.~S., \& {Hlavacek-Larrondo},
  J. 2014{\natexlab{b}}, \mnras, 444, 1236

\bibitem[{{Perley} \& {Butler}(2017)}]{2016arXiv160905940P}
{Perley}, R.~A., \& {Butler}, B.~J. 2017, \apjs, 230, 7

\bibitem[{{Ponman} {et~al.}(1999){Ponman}, {Cannon}, \&
  {Navarro}}]{1999Natur.397..135P}
{Ponman}, T.~J., {Cannon}, D.~B., \& {Navarro}, J.~F. 1999, \nat, 397, 135

\bibitem[{Rafferty {et~al.}(2008)Rafferty, McNamara, \&
  Nulsen}]{2008ApJ...687..899R}
Rafferty, D.~A., McNamara, B.~R., \& Nulsen, P. E.~J. 2008, \apj, 687, 899

\bibitem[{{Sanderson} {et~al.}(2006){Sanderson}, {Ponman}, \&
  {O'Sullivan}}]{2006MNRAS.372.1496S}
{Sanderson}, A.~J.~R., {Ponman}, T.~J., \& {O'Sullivan}, E. 2006, \mnras, 372,
  1496

\bibitem[{Scaife \& Heald(2012)}]{2012MNRAS.423L..30S}
Scaife, A. M.~M., \& Heald, G.~H. 2012, \mnras, 423, L30

\bibitem[{{Schellenberger} \& {Reiprich}(2015)}]{2015arXiv151003708S}
{Schellenberger}, G., \& {Reiprich}, T.~H. 2015, \aap, 583, L2

\bibitem[{{Schellenberger} \& {Reiprich}(2017)}]{2017arXiv170505842S}
---. 2017, \mnras, 469, 3738

\bibitem[{{Slee} {et~al.}(2001){Slee}, {Roy}, {Murgia}, {Andernach}, \&
  {Ehle}}]{2001AJ....122.1172S}
{Slee}, O.~B., {Roy}, A.~L., {Murgia}, M., {Andernach}, H., \& {Ehle}, M. 2001,
  \aj, 122, 1172

\bibitem[{{Sun} {et~al.}(2003){Sun}, {Forman}, {Vikhlinin}, {Hornstrup},
  {Jones}, \& {Murray}}]{2003ApJ...598..250S}
{Sun}, M., {Forman}, W., {Vikhlinin}, A., {et~al.} 2003, \apj, 598, 250

\bibitem[{{Sun} {et~al.}(2006){Sun}, {Jones}, {Forman}, {Nulsen}, {Donahue}, \&
  {Voit}}]{2006ApJ...637L..81S}
{Sun}, M., {Jones}, C., {Forman}, W., {et~al.} 2006, \apjl, 637, L81

\bibitem[{{Sun} {et~al.}(2009){Sun}, {Voit}, {Donahue}, {Jones}, {Forman}, \&
  {Vikhlinin}}]{2009ApJ...693.1142S}
{Sun}, M., {Voit}, G.~M., {Donahue}, M., {et~al.} 2009, \apj, 693, 1142

\bibitem[{{Tozzi} \& {Norman}(2001)}]{2001ApJ...546...63T}
{Tozzi}, P., \& {Norman}, C. 2001, \apj, 546, 63

\bibitem[{{Voit} {et~al.}(2015){Voit}, {Donahue}, {Bryan}, \&
  {McDonald}}]{2015Natur.519..203V}
{Voit}, G.~M., {Donahue}, M., {Bryan}, G.~L., \& {McDonald}, M. 2015, \nat,
  519, 203

\bibitem[{{Wayth} {et~al.}(2015){Wayth}, {Lenc}, {Bell}, {Callingham},
  {Dwarakanath}, {Franzen}, {For}, {Gaensler}, {Hancock}, {Hindson},
  {Hurley-Walker}, {Jackson}, {Johnston-Hollitt}, {Kapi{\'n}ska}, {McKinley},
  {Morgan}, {Offringa}, {Procopio}, {Staveley-Smith}, {Wu}, {Zheng}, {Trott},
  {Bernardi}, {Bowman}, {Briggs}, {Cappallo}, {Corey}, {Deshpande}, {Emrich},
  {Goeke}, {Greenhill}, {Hazelton}, {Kaplan}, {Kasper}, {Kratzenberg},
  {Lonsdale}, {Lynch}, {McWhirter}, {Mitchell}, {Morales}, {Morgan}, {Oberoi},
  {Ord}, {Prabu}, {Rogers}, {Roshi}, {Shankar}, {Srivani}, {Subrahmanyan},
  {Tingay}, {Waterson}, {Webster}, {Whitney}, {Williams}, \&
  {Williams}}]{2015PASA...32...25W}
{Wayth}, R.~B., {Lenc}, E., {Bell}, M.~E., {et~al.} 2015, \pasa, 32, e025

\bibitem[{Whiting(2012)}]{2012MNRAS.421.3242W}
Whiting, M.~T. 2012, \mnras, 421, 3242

\bibitem[{{Willingale} {et~al.}(2013){Willingale}, {Starling}, {Beardmore},
  {Tanvir}, \& {O'Brien}}]{2013MNRAS.tmp..859W}
{Willingale}, R., {Starling}, R.~L.~C., {Beardmore}, A.~P., {Tanvir}, N.~R., \&
  {O'Brien}, P.~T. 2013, \mnras, 431, 394

\bibitem[{{Zahid} {et~al.}(2016){Zahid}, {Geller}, {Fabricant}, \&
  {Hwang}}]{2016arXiv160704275Z}
{Zahid}, H.~J., {Geller}, M.~J., {Fabricant}, D.~G., \& {Hwang}, H.~S. 2016,
  \apj, 832, 203

\bibitem[{ZuHone {et~al.}(2010)ZuHone, Markevitch, \&
  Johnson}]{2010ApJ...717..908Z}
ZuHone, J.~A., Markevitch, M., \& Johnson, R.~E. 2010, \apj, 717, 908

\end{thebibliography}

	\appendix
	\section{Temperature and abundance maps}
	\label{app:maps}

\begin{figure}
	\centering
	\resizebox{0.99\hsize}{!}{\includegraphics[width=0.47\textwidth]{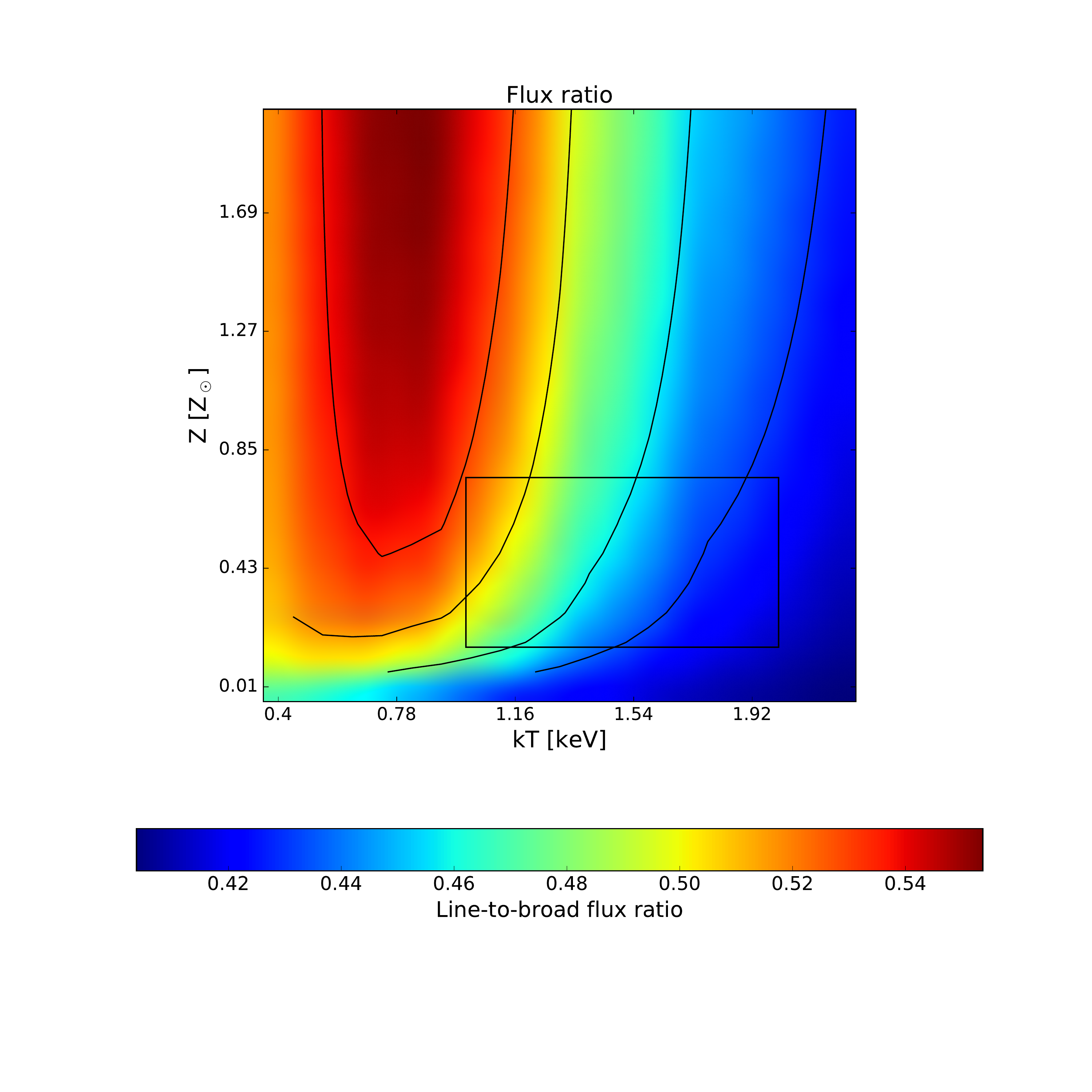} 	\includegraphics[width=0.47\textwidth]{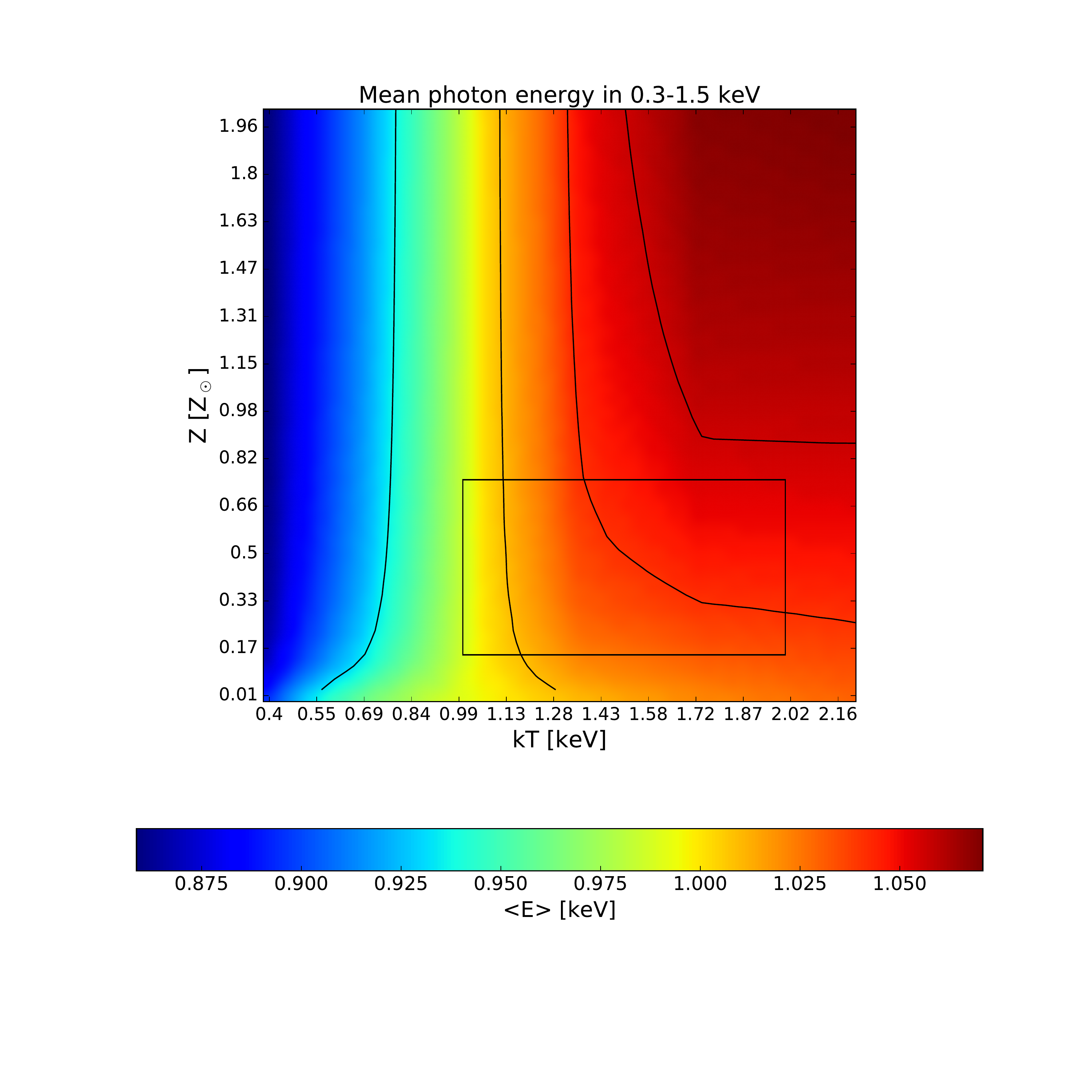}}
	\caption{Simulations to connect the two observables with ICM temperature and metallicity. Both images show one slice of the 3D cubes (for the maximum source-to-background ratio). The black box shows an important range of the two parameters.}
	\label{fig:simu}
\end{figure}

Spatial variations of temperature and abundances of heavy elements in the ICM are crucial to investigate the merging history of a galaxy cluster. Apart from radial profiles the information provided by detailed maps is of great importance because merging processes usually create an asymmetric distribution of the temperature and abundances. 
The ability to constrain with reasonable accuracy the temperature and abundance from a spectral fit depends strongly on the number of counts, but also on the temperature of the system. For low temperature galaxy groups one needs several hundreds of counts to constrain the temperature and even more for a reliable abundance measurement. With available exposure times, this does not allow the detailed investigation of NGC741 by a temperature or abundance map, especially not in lower surface brightness regions. 

We present a method to create temperature and abundance maps for low temperature galaxy groups (up to $\SI{2}{keV}$) by combining the information from images of the system in three different bands: It has already been shown (\citealp{2009ApJ...705..624D}) that the mean energy of photons in a soft band covering the Fe-L line complex is a good measure for the ICM temperature and fairly independent of the abundance of heavy elements. 
The flux of the Fe-L line itself strongly depends on the abundance. By measuring the flux of the Fe-L line with respect to a broad band flux an observable for the abundance can be constructed. This observable still depends on the gas temperature, but this influence can be almost completely eliminated by combining the new observable with the mean energy. 

Our simulations show that there exists a clear, almost abundance independent, relation between the mean photon energy and the ICM temperature. Furthermore, the ratio of the Fe-L line flux to broad band flux is sensitive to the abundance. But the two observables, the line-to-broad-band flux ratio and the mean photon energy have both dependencies on the ICM temperature and abundance, and also they depend on the source to background ratio. 

We simulate spectra with an absorbed \verb|apec| model (redshift and $N_\mathrm{H}$ frozen to cluster values) and give 20 uniformly distributed temperatures, $\SIrange{0.4}{2.2}{keV}$, 20 different abundances, $\SIrange{0.01}{2.0}{Z_\odot}$, and 10 different source to background ratios, $0.5 - 140$, to construct a 3D matrix for each observable. 
These matrices are linearly interpolated to obtain the most likely values of temperature and abundance at each pixel of the input image. Figure \ref{fig:simu} shows the one slice of the 3D matrices at maximum source-to-background ratio.

For the simulations we use the same responses created for the actual observations to account for the low energy degrading of the Chandra effective area. For the combination of more than one observation we weight them in the simulations according to their actual exposure times. 

The actual images are produced in the the $\SIrange{0.3}{4}{keV}$ band for the broad flux, the $\SIrange{0.7}{1.3}{keV}$ band for the Fe-L line flux, and the $\SIrange{0.3}{1.5}{keV}$ band for the mean energy to estimate the gas temperature. For our purposes these bands turned out to be well suited. 
Especially the band to determine the mean photon energy has to be extended toward higher energies to also allow a temperature determination up to $\SI{2}{keV}$.
To have comparable statistics in each spatial bin, we create a mask using the adaptive binning task \verb|dmnautilus| and we require a signal-to-noise of at least 10 in the the Fe-L line band. Point sources are detected via the ciao task \verb|wavdetect| and excluded for the analysis. It turns out that strong unremoved point sources like AGNs show up in the abundance map as regions with very low abundance.

We tested our results for the influence of multi temperature components. As suggested in \cite{2000MNRAS.311..176B}, the Iron bias effect produces low-biased abundance measurements in multi temperature spectra which are fitted with just a single component. We estimated if this effect is also present in our method to produce an abundance map.
We find that in general the maps are affected by the Iron bias in a very similar way as the spectral fit of a single temperature component. For example, for $\SI{0.4}{keV}$ and $\SI{1}{keV}$ gas with a emission measure ratio (cold-to-hot, EMR) of 1, we find a 60\% lower abundance for both methods. If we change the EMR to 0.1 (more hotter gas), we find a lower abundance of 12\%  for the spectral fit, and 20\% for the map. Instead if we have 10 times more colder gas than hotter, we find 44\% lower abundances for the map and 52\% for the spectral fit. If we increase the temperature of the cold component, we also find that the difference of the map abundance to the real one is smaller.
In high quality data an indication for multi phase gas is a bad spectral fit, e.g., $\chi^2 >> 1$. Since we have no measure of the goodness of fit for a thermal model available in our temperature and abundance maps, we cannot use this quantity to alleviate the influence of the iron bias.

\end{document}